\documentclass[aps,prb,twocolumn,reprint,groupedaddress]{revtex4}
\usepackage{graphicx}
\usepackage{amsmath}
\usepackage{bbold}
\usepackage[usenames, dvipsnames]{color}

\begin{document}

\title{ Measurements of magnetization on the Sierpi\'{n}ski carpet }

\author{Jozef \textsc{Genzor}$^{1, 2, 3}$}
\author{Andrej {\textsc Gendiar}$^{4}$}
\author{Tomotoshi {\textsc Nishino}$^{1}$}
\affiliation{$^1$Department of Physics, Graduate School of Science, Kobe University, Kobe 657-8501, Japan}
\affiliation{$^2$Physics Division, National Center for Theoretical Sciences, Taipei 10617, Taiwan}
\affiliation{{$^3$Physics Division, National Center for Theoretical Sciences, National Taiwan University, Taipei 10617, Taiwan}}
\affiliation{$^4$Institute of Physics, Slovak Academy of Sciences, D\'ubravsk\'a cesta 9, SK-845 11, Bratislava, Slovakia}

\date{\today}

\begin{abstract}
Phase transition of the classical Ising model on the Sierpi\'{n}ski carpet, which has the fractal dimension $\log_3^{~} 8 \approx 1.8927$, is studied by an adapted variant of the higher-order tensor renormalization group method. 
The second-order phase transition is observed at the critical temperature $T_{\rm c}^{~} \approx 1.478$.
Position dependence of local functions is studied through impurity tensors inserted at different locations on the fractal lattice. 
The critical exponent $\beta$ associated with the local magnetization varies by two orders of magnitude, depending on lattice locations, whereas $T_{\rm c}^{~}$ is not affected.
%
Furthermore, we employ automatic differentiation to accurately and efficiently compute the average spontaneous magnetization per site as a first derivative of free energy with respect to the external field, yielding the global critical exponent of $\beta \approx 0.135$.
%
\end{abstract}

\maketitle

\section{Introduction}

Understanding of phase transitions and critical phenomena plays an important role in the condensed matter 
physics\cite{phase_trans}. Systems on regular lattices are of the major target of such studies, where elementary 
models exhibit translationally invariant states, which are scale invariant at criticality. It has been known 
that critical behavior is controlled by global properties, such as dimensionality and symmetries. 
This is the concept of the \textit{universality}. 

If we focus our attention on inhomogeneous lattices, there is a group of fractal lattices, which 
are self-similar and exhibit non-integer Hausdorff dimensions. Geometrical details, such as lacunarity 
and connectivity, could thus modify the properties of their critical phenomena. An important aspect of the fractal lattices 
is the ramification, which is the smallest number of bonds that have to be cut in order to isolate an 
arbitrarily large bounded subset surrounding a point. In the early studies by Gefen 
{\it et al.}\cite{Gefen1, Gefen2, Gefen3, Gefen4}, it was shown that the short-range classical spin 
models on finitely ramified lattices exhibit no phase transition at nonzero temperature. 

The ferromagnetic Ising model on the fractal lattice that corresponds to the Sierpi\'{n}ski carpet is one of the most extensively studied models with fractal lattice geometry. 
Monte Carlo studies combined with the finite-size scaling method have been performed\cite{Carmona, Monceau1, Monceau2, Pruessner, Bab, Bab2}, including Monte Carlo renormalization group method\cite{MCRG}. 
The critical temperature $T_{\rm c}^{~}$ is relatively well estimated within the narrow range $1.47 \lesssim T_{\rm c}^{~} \lesssim 1.50$, where one of the most recent estimate is {$T_{\rm c}^{~} = 1.495(5)$ by Bab {\it et al.}\cite{Bab2}}. 
On the other hand, estimates of critical exponents are still fluctuating, since it is rather hard to collect sufficient numerical data for a precise finite-size scaling analysis\cite{RSRG}. 
This is partially so because an elementary lattice unit can contain too many sites, and there is a variety of choices with respect to boundary conditions. 
This situation persists even in a recent study by means of a path-counting approach\cite{Perreau}. 
Yet, a number of issues remains unresolved concerning uniformity of fractal systems in the thermodynamic limit\cite{Bab}.

Recently, we show that the higher-order tensor renormalization group (HOTRG) method\cite{HOTRG} 
can be used as an appropriate numerical tool for studies of certain types of fractal systems {\cite{2dising, APS, gasket, j1j2}}.
The method is based on the real-space renormalization group, and, therefore, the self-similar property of fractal 
lattices can be treated in a natural manner. In this article, we apply the HOTRG method to the Ising model 
on the fractal lattice that corresponds to the Sierpi\'{n}ski carpet. The method enables us to estimate 
$T_{\rm c}^{~}$ from the temperature dependence of the entanglement entropy $s( T )$. In order to 
check the uniformity in the thermodynamic functions, we choose three distinct locations on the lattice, 
and calculate the local magnetization $m( T )$ and the bond energy $u( T )$. As it is trivially expected, 
these local functions, $m( T )$ and $u( T )$, yield the identical $T_{\rm c}^{~}$. Contrary to the naive intuition,
the critical exponent $\beta$, which is associated with the local magnetization $m( T )\propto(T_{\rm c}^{~}-T)^{\beta}$, 
strongly depends on the location of measurement, and the estimated exponent $\beta$ can vary within two orders 
of magnitude with respect to the three different locations on the fractal lattice, where the local functions are calculated.

%
Recent research has demonstrated the effectiveness of automatic differentiation, a technique derived from deep learning, for accurately and efficiently computing higher-order derivatives in tensor network algorithms~\cite{ad1, ad2}. 
Automatic differentiation is based on a computation graph representing the sequence of elementary computation steps in a directed acyclic graph. 
This technology can propagate gradients with machine precision throughout the computation process. 
In tensor network algorithms, the implementation of numerically stable differentiation through linear algebra operations, such as the Singular Value Decomposition (SVD), is crucial. 
By applying automatic differentiation to our tensor network fractal, we can accurately calculate the average spontaneous magnetization as the first derivative of the free energy with respect to the external field. 
Unlike numerical derivatives, this approach avoids introducing numerical errors due to finite step sizes. 
Once the magnetization is computed, we can extract the global critical exponent $\beta$.
%

Structure of this article is as follows. In the next section, we explain the recursive construction of 
the fractal lattice, and express the partition function of the system in terms of contractions among 
tensors. In Sec.~III we introduce HOTRG method for the purpose of keeping the numerical cost 
realistic. The way of measuring the local functions $m( T )$ and $u( T )$ is explained. Numerical results 
are shown in Sec.~IV. Position dependence on the local functions is observed. In the last section, we 
summarize the obtained results, and discuss the reason for the pathological behavior of the fractal system.

\section{Model representation}  

\begin{figure}[tb]
\includegraphics[width=0.48\textwidth]{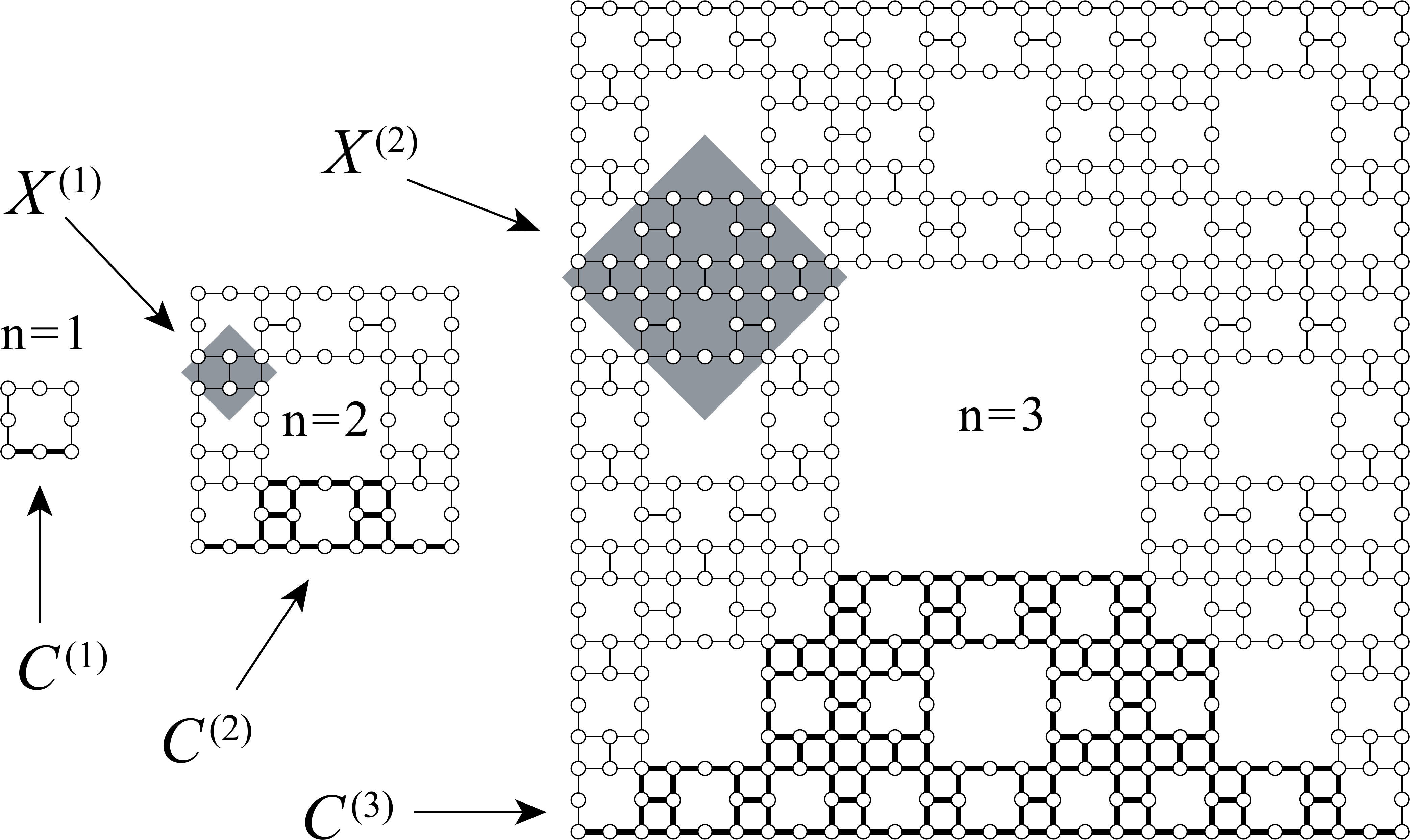}
\caption{
Build-up process of a discrete analog of the Sierpi\'{n}ski carpet. The circles represent the lattice points, 
where the Ising spins are located. The vertical and horizontal links denote the interacting pairs. The first three 
units $n = 1$, $2$, and $3$ are shown. For each unit $n$, we draw the corners $C^{(n)}_{~}$ by the
thick lines. We label the shaded regions $X^{(1)}_{~}$ and $X^{(2)}_{~}$.}

\label{fig:Fig_1}
\end{figure}

There are several different types of discrete lattices that can be identified as the Sierpi\'{n}ski carpet. 
Among them, we choose the one constructed by the extension process shown in Fig.~\ref{fig:Fig_1}.
In the first step ($n = 1$), there are eight spins in the unit, as it is shown on the left. The Ising spins 
$\sigma = \pm 1$ are represented by the circles, and the ferromagnetic nearest-neighbor interactions 
are denoted by the horizontal and vertical lines. In the second step ($n = 2$), the eight units are grouped 
to form a new extended unit, as shown in the middle. Now, there are $64$ spins on the $9 \times 9$ square
lattice grid. On the right side, we show the third step ($n = 3$). Generally, in the $n$-th step, an extended unit 
contains $8^n_{~}$ spins on the $3^n_{~} \times 3^n_{~}$ lattice. The Hausdorff dimension of this lattice 
is $d_{\rm H}^{~} = \log_3^{~} 8 \approx 1.8927$ in the thermodynamic limit $n\to\infty$.

In the series of the extended units we have thus constructed, there is another type of the recursive structure.
In Fig.~\ref{fig:Fig_1} at the bottom of each unit, we have drawn a pyramid-like area by the thick lines.
One can identify four such pyramid-like areas within each unit (enumerated by $n$), and each area can be called
the {\it corner} $C^{(n)}_{~}$. The corners are labeled $C^{(1)}_{~}$, $C^{(2)}_{~}$, and $C^{(3)}_{~}$ from
left to right therein. It should be noted that there are only 
$2^{n-1}_{~}$ spin sites in common, where two adjacent corners meet.

In the case $n = 2$ drawn in the middle, we shaded a region on the left, which contains six sites, and label
the region $X^{(1)}_{~}$. Having observed the corner $C^{(2)}_{~}$ at the bottom, we found out that the corner
consists of two rotated pieces of $X^{(1)}_{~}$ and the four pieces of $C^{(1)}_{~}$. In $n = 3$, we shaded
a larger region $X^{(2)}_{~}$ (in the similar manner as $X^{(1)}_{~}$), which now contains 36 sites.
We can recognize that $X^{(2)}_{~}$ consists of seven pieces of $X^{(1)}_{~}$ and the two pieces of $C^{(1)}_{~}$.
We have thus identified the following recursive relations, which build up the fractal:
\begin{itemize}
\item{} Each $n$-th unit contains 4 pieces of $C^{(n)}_{~}$,
\item{} $\!C^{(n+1)}_{~}\!$ contains 2 pieces of $X^{(n)}_{~}\!$ and 4 pieces of $C^{(n)}_{~}\!$,
\item{} $\!X^{(n+1)}_{~}\!$ contains 7 pieces of $X^{(n)}_{~}\!$ and 2 pieces of $C^{(n)}_{~}\!$.
\end{itemize}

\begin{figure}[tb]
\includegraphics[width=0.32 \textwidth]{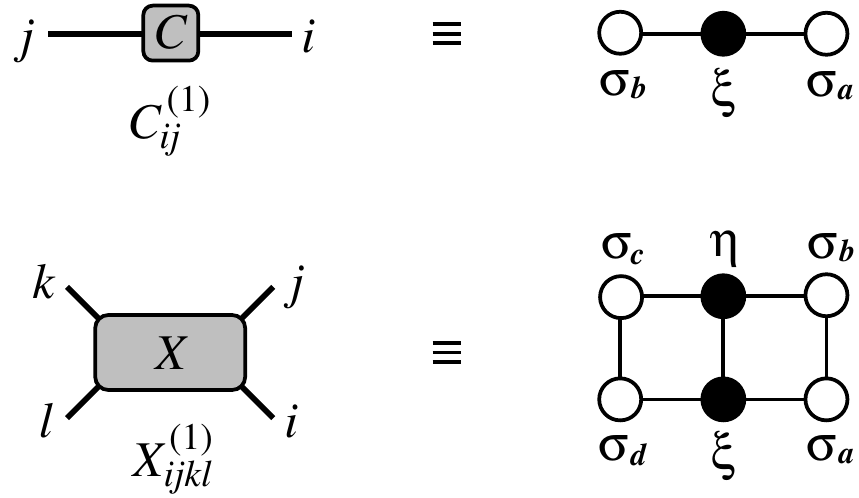}
\caption{
Structure of the initial corner matrix $C^{(1)}_{ij}$ in Eq.~\eqref{Eq_3} and 
the 4-leg tensor $X^{(1)}_{ijkl}$ in Eq.~\eqref{Eq_6}.}
\label{fig:Fig_2}
\end{figure}

The Hamiltonian of the Ising model, which is constructed on the series of finite-size systems $n = 1, 2, 3, \cdots$,
has the form
\begin{equation}
H^{(n)}_{~} = - J \sum_{\left< a b \right>} \sigma_a^{~} \sigma_b^{~} \, .
\label{Eq_1}
\end{equation}
The summation runs over all pairs of the nearest-neighbor Ising spins, as shown by the circles in
Fig.~\ref{fig:Fig_1}. The spin positions are labeled by the lattice indices $a$ and $b$. They are connected by
the lines, which correspond to the ferromagnetic interaction $J > 0$, and no external magnetic field is imposed.
First we calculate the partition function (expressed in arbitrary step $n$)
\begin{equation}
Z^{(n)}_{~} = \sum \exp\biggl[ -  \frac{~ H^{(n)}_{~} }{ {k_{\rm B}^{~}T} } \biggr]
\label{Eq_2}
\end{equation}
as a function of temperature $T$, where the summation is taken over all spin configurations, 
and where $k_{\rm B}^{~}$ denotes the Boltzmann constant. At initial step $n=1$, we define the {\it corner} matrix 
\begin{equation} 
C_{ij}^{(1)} = \sum\limits_{\xi = \pm1}^{~} 
\exp\bigl[ K \xi \left(\sigma_a^{~} + \sigma_b^{~} \right) \bigr] \, ,
\label{Eq_3}
\end{equation}
where $K = J / k_{\rm B}^{~} T$, and the matrix indices $i = ( \sigma_a^{~} + 1 ) / 2$ and 
$j = ( \sigma_b^{~} + 1 ) / 2$ take the value either $0$ or $1$. The structure on the right-hand side is 
graphically shown in Fig.~\ref{fig:Fig_2} (top), and the summation taken over the spin $\xi$ is denoted by the filled 
circle. We have chosen the ordering of the indices $i$ and $j$, which is opposite if comparing $C_{ij}^{(1)}$ 
with the corresponding graph. The partition function of the smallest unit ($n = 1$), which contains 
8-spins, is then expressed as 
\begin{equation}
Z^{(1)}_{~} = \sum_{ijkl}^{~} C_{ij}^{(1)} \, C_{jk}^{(1)} \, C_{kl}^{(1)} \, C_{li}^{(1)} \, ,
\label{Eq_4}
\end{equation}
and can be abbreviated to ${\rm Tr} \, \bigl[ C^{(1)}_{~} \bigr]^4_{~}$. 
We will express $Z^{(n)}_{~}$ for arbitrary $n > 1$ in the same trace form
\begin{equation}
Z^{(n)}_{~} = {\rm Tr} \, \bigl[ C^{(n)}_{~} \bigr]^4_{~}
\label{Eq_5}
\end{equation}
by means of the corner matrix $C^{(n)}_{ij}$, where each one undergoes {\it extensions}, as we define in the following.

\begin{figure}[tb]
\includegraphics[width=0.43 \textwidth]{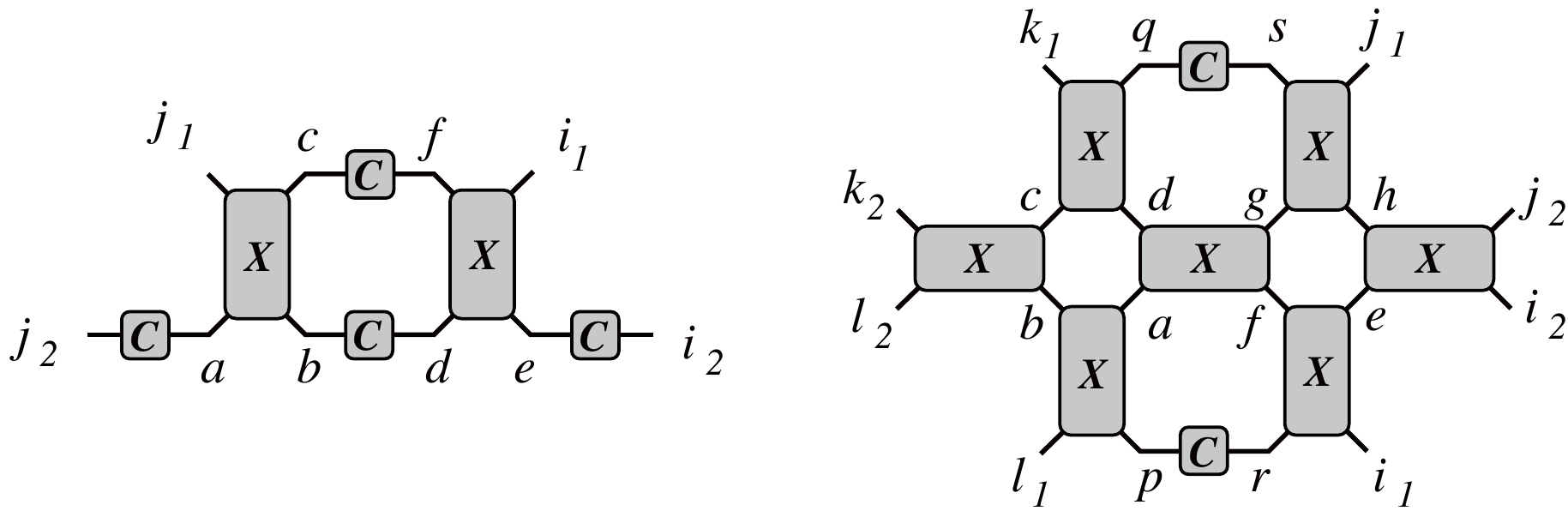}
\caption{
Extension of the local matrix $C^{(n)}_{~}$ in Eq.~\eqref{Eq_7} (on the left) and the tensor $X^{(n)}_{~}$ in 
Eq.~\eqref{Eq_8} (on the right).}
\label{fig:Fig_3}
\end{figure}

Let us notice that the region $X^{(1)}_{~}$ appears from the step $n = 2$. The Boltzmann weight
corresponding to this region $X^{(1)}_{~}$ can be expressed by the 4-leg (order-4) tensor
\begin{equation}
\begin{split}
X_{ijkl}^{(1)} = \sum_{\xi \eta}^{~} & \exp\bigl[ K \left( \sigma_a^{~} \sigma_b^{~} + 
\sigma_c^{~} \sigma_d^{~} + \xi \eta \right)  \bigr] \\
\times & \exp \bigl[ K \xi \left( \sigma_a^{~} + \sigma_d^{~} \right) + 
K \eta \left( \sigma_b^{~} + \sigma_c^{~} \right) \bigr] \, ,
\end{split}
\label{Eq_6}
\end{equation}
where the spin locations are depicted in Fig.~\ref{fig:Fig_2} (bottom). We have additionally introduced 
new indices $k = ( \sigma_c^{~} + 1 ) / 2$ and $l = ( \sigma_d^{~} + 1 ) / 2$. 
Now we can mathematically represent the recursive relations in terms of contractions among the 
matrices $C^{(n)}_{~}$ and tensors $X^{(n)}_{~}$. Figure~\ref{fig:Fig_3} shows the graphical representation 
of the extension processes.  Taking the contraction among the two tensors $X^{(n)}_{~}$ and the 
four matrices $C^{(n)}_{~}$, as shown in Fig.~\ref{fig:Fig_3} (left), we obtain the {\it extended} corner 
matrix $C^{(n+1)}_{~}$ through the corresponding formula
\begin{equation}
\begin{split}
C_{ij}^{(n+1)} &= C_{( i_1^{~} i_2^{~}) (j_1^{~} j_2^{~})}^{(n+1)} \\ 
& = \sum_{a b c d e f}^{~} 
C_{a j_2^{~}}^{(n)} X_{a b c j_1^{~}}^{(n)} C_{f c}^{(n)} C_{d b}^{(n)} X_{d e i_1^{~} f}^{(n)} C_{i_2^{~} e}^{(n)} \, ,
\end{split} 
\label{Eq_7}
\end{equation}
where the new indices $i$ and $j$, respectively, represent the grouped indices $( i_1^{~} i_2^{~})$ and
$(j_1^{~} j_2^{~})$. Apparently, the diagram in Fig.~\ref{fig:Fig_3} (left) is more convenient than Eq.~\eqref{Eq_7}
for the better understanding of the contraction geometry. This relation can be easily checked for the case 
$n = 1$ after comparing Figs.~\ref{fig:Fig_1}, \ref{fig:Fig_2}, and \ref{fig:Fig_3}. 

Similarly, the extension process from $X^{(n)}_{~}$ to $X^{(n+1)}_{~}$ shown in Fig.~\ref{fig:Fig_3} (right)
can be expressed by the formula
\begin{equation}
\begin{split}
X_{ijkl}^{(n+1)} & = X_{( i_1^{~} i_2^{~}) (j_1^{~} j_2^{~}) ( k_1^{~} k_2^{~}) (l_1^{~} l_2^{~})}^{(n+1)} \\
& =\sum\limits_{{a b c d e f}\atop{g h p r q s}} X_{a b l_1^{~} p}^{(n)} X_{b c k_2^{~} l_2^{~}}^{(n)} 
X_{c d q k_1^{~}}^{(n)} X_{f g d a}^{(n)} \\
& \hspace{1.4cm} X_{e f r i_1^{~}}^{(n)} X_{g h j_1^{~} s}^{(n)}
X_{i_2^{~} j_2^{~} h e}^{(n)} C_{r p}^{(n)} C_{s q}^{(n)} \, , \end{split} 
\label{Eq_8}
\end{equation}
where we have again abbreviated the grouped indices to $i = (i_1^{~} i_2^{~})$, $j = (j_1^{~} j_2^{~})$, 
$k = (k_1^{~} k_2^{~})$, and $l = (l_1^{~} l_2^{~})$. This relation can be checked for the case $n = 1$ by
comparing the area $X^{(1)}_{~}$ and $X^{(2)}_{~}$ in Fig.~1.

Through the iterative extension of the tensors, we can {\it formally} obtain the corner matrix 
$C^{(n)}_{ij}$ for arbitrary $n$, and express $Z^{(n)}_{~}$ by Eq.~\eqref{Eq_5}. The free energy 
per spin is then
\begin{equation}
f^{(n)}_{~} = - \frac{1}{8^n} \, k_{\rm B}^{~} T \ln Z^{(n)}_{~}
\label{Eq_9}
\end{equation}
since the $n$-th unit contains $8^n_{~}$ spins. This function converges to a value $f^{(\infty)}_{~}$ in the 
thermodynamic limit $n \rightarrow \infty$, where convergence with respect to $n$ is rapid, and 
$n = 35$ is sufficient in the numerical analyses. The specific heat per site can be evaluated by 
taking the second derivative of the free energy
$c_f^{~}( T ) = - T \frac{\partial^2_{~}}{\partial T^2_{~}}f^{(\infty)}_{~}$.
%
Furthermore, the global spontaneous magnetization $m_f$ can be evaluated as the first derivative of the free energy $f^{(\infty)}_{~}$ with respect to the external field $h$
\begin{equation}
m_{f}^{~}(T) = -\left. \dfrac{\partial f^{(\infty)}_{~}}{\partial h} \right|_{h \to 0} \, . 
\label{Eq:glob_mag}
\end{equation}
To avoid numerical errors due to the finite step as in the case of numerical derivative, we calculate the global magnetization $m_f$ according to the Eq.~\eqref{Eq:glob_mag} accurately and efficiently using automatic differentiation applied to the tensor network program for the partition function of the fractal lattice in our study. 
%

\section{Renormalization Group Transformation}

The matrix dimension of $C^{(n)}_{~}$ is $2^{n-1}_{~}$ by definition. Therefore, it is impossible to keep 
all the matrix elements faithfully in numerical analysis, when $n$ is large. The situation
is severer for $X^{(n)}_{~}$, which has four indices. By means of the HOTRG method~\cite{HOTRG}, 
it is possible to reduce the tensor-leg dimension, the degree of freedom, down to a 
realistic number. The reduction process is performed by the renormalization group 
transformation $U$, which is created from the higher-order singular value decomposition 
(SVD)~\cite{hosvd} applied to the extended tensor $X^{(n+1)}_{ijkl}$. 

Suppose that the tensor-leg dimension in $X^{(n)}_{ijkl}$ is $D$ for each index, i.e., $i,j,k,l=0,1,\dots,D-1$.
As we have shown in Eq.~\eqref{Eq_8}, the dimension of the grouped index $i = ( i_1^{~} i_2^{~} )$ in 
$X_{(i_1^{~} i_2^{~}) (j_1^{~} j_2^{~}) (k_1^{~} k_2^{~}) (l_1^{~} l_2^{~}) }^{(n+1)}$ is equal to $D^2_{~}$. 
We reshape the four tensor indices to form a rectangular matrix with the grouped index $(i_1^{~} i_2^{~})$
and the remaining grouped index $(j_1^{~} j_2^{~} k_1^{~} k_2^{~} l_1^{~} l_2^{~})$ with the dimension $D^6$.
Applying the singular value decomposition to the reshaped tensor, we obtain
\begin{equation}
X_{(i_1^{~} i_2^{~}) (j_1^{~} j_2^{~} k_1^{~} k_2^{~} l_1^{~} l_2^{~}) }^{(n+1)} = 
\sum_{\xi}^{~} U_{(i_1^{~} i_2^{~}) \, \xi}^{~} \, \omega_{\xi}^{~} \, 
V^{~}_{(j_1^{~} j_2^{~} k_1^{~} k_2^{~} l_1^{~} l_2^{~}) \, \xi } \, ,
\label{Eq_10}
\end{equation}
where $U$ and $V$ are generalized unitary, i.e. orthonormal, matrices $U^T_{~}U=V^T_{~}V=\mathbb{1}$. We assume the decreasing 
order for the singular values $\omega_{\xi}^{~}$ by convention. Keeping $D$ dominant degrees of 
freedom for the index $\xi$ at most, we regard the matrix $U_{(i_1^{~} i_2^{~}) \, \xi}^{~}$ as the renormalization 
group (RG) transformation from $(i_1^{~} i_2^{~})$ to the renormalized index $\xi$. 
For the purpose of clarifying the relation between the original pair of indices $(i_1^{~} i_2^{~})$ and 
the renormalized index $\xi$, we {\it rename} $\xi$ to $i$ and write the RG transformation as
$U_{(i_1^{~} i_2^{~}) \, i}^{~}$. In the same manner, we obtain $U_{(j_1^{~} j_2^{~}) \, j}^{~}$, 
$U_{(k_1^{~} k_2^{~}) \, k}^{~}$, and $U_{(l_1^{~} l_2^{~}) \, l}^{~}$, where we have distinguished
the transformation matrices by their indices. 

\begin{figure}[tb]
\includegraphics[width=0.45 \textwidth]{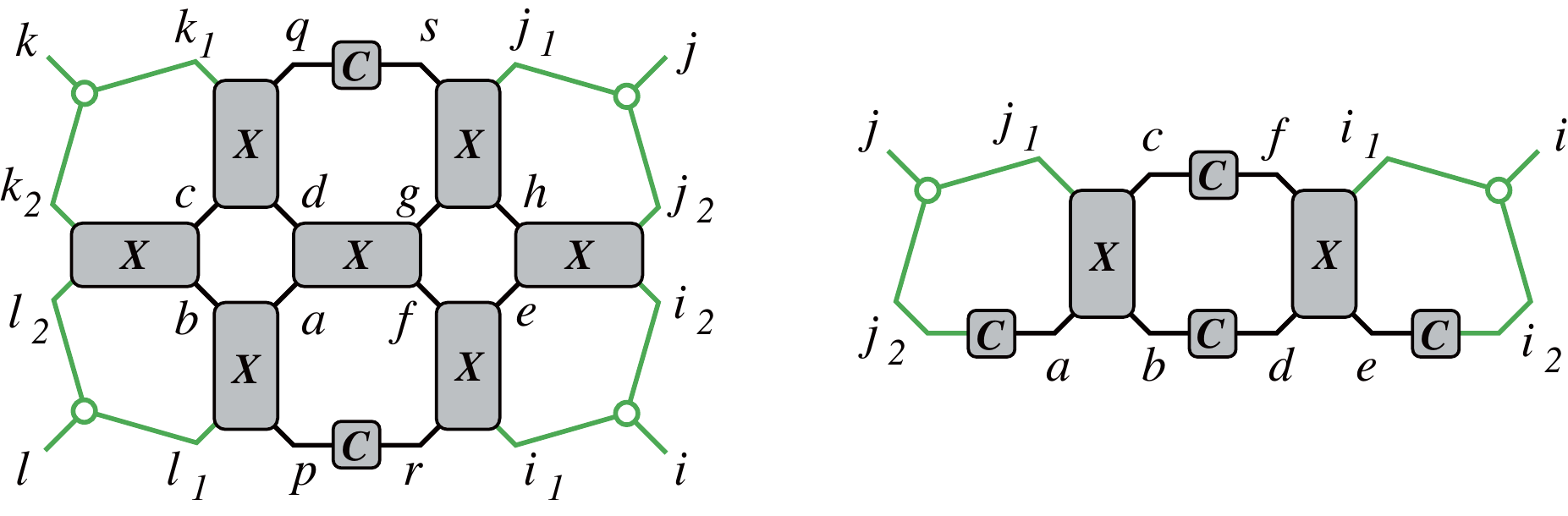}
\caption{(Color online) 
The renormalization group transformations in Eq.~\eqref{Eq_11} (on the left) and Eq.~\eqref{Eq_12} (on the right)
applied, respectively, to Eq.~\eqref{Eq_8} and \eqref{Eq_7} (cf. Fig.~\ref{fig:Fig_3}).
}
\label{fig:Fig_4}
\end{figure}

The RG transformation is then performed as
\begin{align}
\begin{split}
X_{i j k l }^{(n+1)}  \leftarrow \sum_{{i_1 i_2 j_1 j_2}\atop{k_1 k_2 l_1 l_2}}
U_{(i_1^{~} i_2^{~}) \, i}^{~} & \, U_{(j_1^{~} j_2^{~}) \, j}^{~} \, 
U_{(k_1^{~} k_2^{~}) \, k}^{~} \, U_{(l_1^{~} l_2^{~}) \, l}^{~} \\
 & X_{(i_1^{~} i_2^{~}) (j_1^{~} j_2^{~}) (k_1^{~} k_2^{~}) (l_1^{~} l_2^{~}) }^{(n+1)} \, , 
\label{Eq_11}
\end{split}
\end{align}
where the sum is taken over the indices on the connected lines in Fig.~\ref{fig:Fig_4} (left). 
The left arrow used in Eq.~\eqref{Eq_11} represents the replacement of the expanded tensor
$X_{(i_1^{~} i_2^{~}) (j_1^{~} j_2^{~}) (k_1^{~} k_2^{~}) (l_1^{~} l_2^{~}) }^{(n+1)}$
for the renormalized one $X_{i j k l }^{(n+1)}$.
Since the RG transformation matrices $U$ are obtained from SVD applied to 
$X_{(i_1^{~} i_2^{~}) (j_1^{~} j_2^{~}) (k_1^{~} k_2^{~}) (l_1^{~} l_2^{~}) }^{(n+1)}$, 
there is no guarantee that the RG transformation can be straightforwardly applied to 
$C_{(i_1^{~} i_2^{~}) (j_1^{~} j_2^{~})}^{(n+1)}$, as we have defined in Eq.~\eqref{Eq_7}. It has been 
confirmed that the transformation
\begin{equation}
C_{i j}^{(n+1)} \leftarrow \sum\limits_{{i_1^{~} i_2^{~}}\atop{j_1^{~} j_2^{~}}} U_{(i_1^{~} i_2^{~}) \, i}^{~} \, U_{(j_1^{~} j_2^{~}) \, j}^{~}  \, 
C_{(i_1^{~} i_2^{~}) (j_1^{~} j_2^{~})}^{(n+1)}
\label{Eq_12}
\end{equation}
is of use in the actual numerical calculation. The corresponding diagram is shown in Fig.~\ref{fig:Fig_4} (right).

We add a remark on the choice of the transformation matrix $U$. In a trial calculation, once we tried to 
create $U$ from the corner matrix $C_{i j}^{(n+1)}$ by both SVD and diagonalization. However, we encountered numerical 
instabilities, in which the singular values (or eigenvalues) decayed to zero too rapidly, especially, when $n$ was 
large. Thus, we always create $U$ from SVD that is applied to $X^{(n+1)}_{ijkl}$ only.

With the use of these RG transformations, it is possible to repeat the extension processes 
in Eq.~\eqref{Eq_7} and \eqref{Eq_8}, and to obtain a good numerical estimate for $Z^{(n)}_{~}$ and $f^{(n)}_{~}$ in
Eq.~\eqref{Eq_9}. The actual numerical calculations in this work were performed by a slightly modified 
procedure, which we describe in detail in Appendix~\ref{app_A}. 
We split $X^{(n)}_{ijkl}$ into two halves and represent each part by $3$-leg tensor. 
This computational trick allowed us to increase the leg-dimension up to $D = 28$, or even larger. 

\subsection{Impurity tensors} 

\begin{figure}[tb]
\includegraphics[width=0.40 \textwidth]{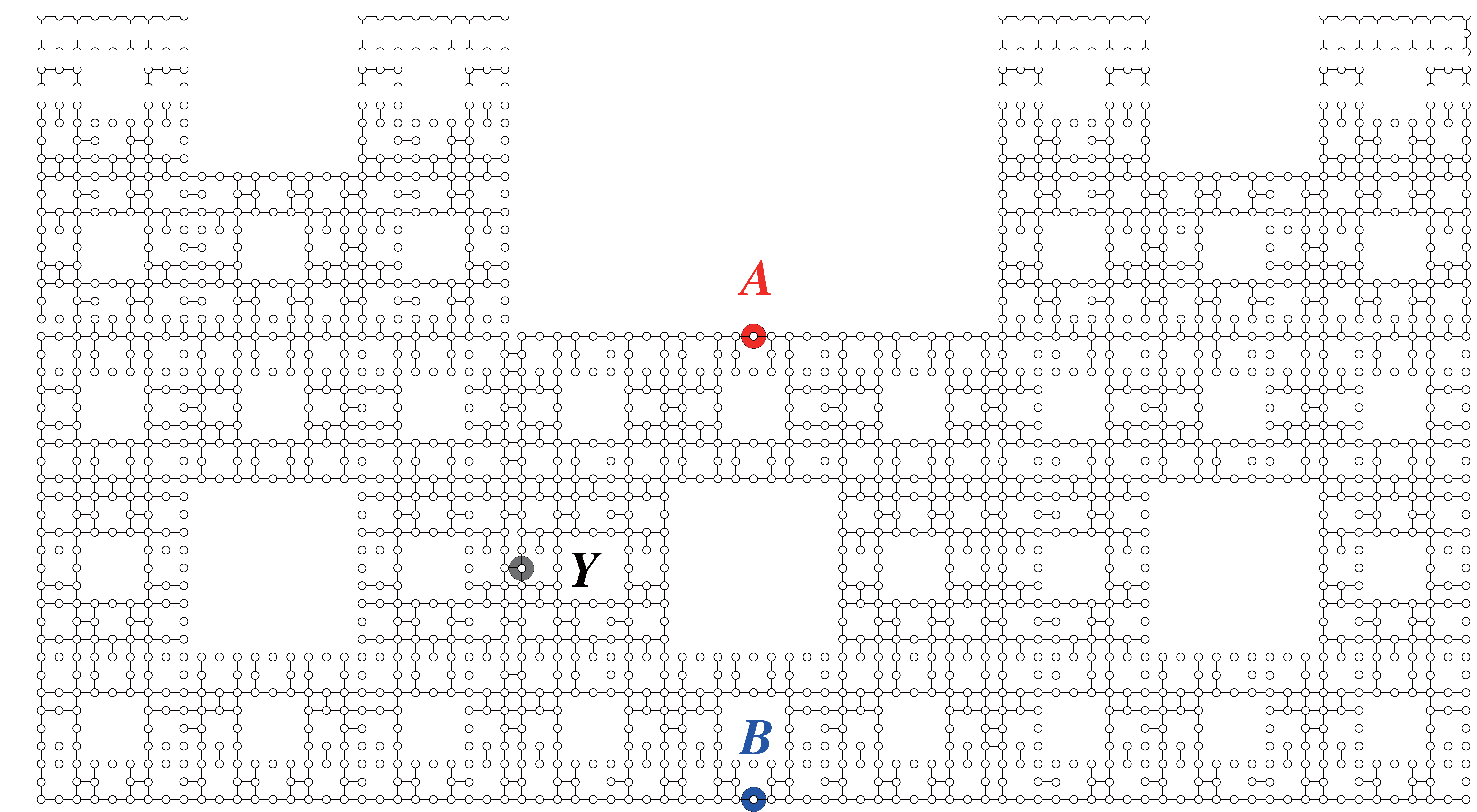}
\caption{(Color online) 
Three positions A (on inner boundary), B (on outer boundary), and Y (in innermost position) chosen for the
observation of local functions. The lower half of the unit $n = 4$ is drawn only.}
\label{fig:Fig_5}
\end{figure}

In the framework of the HOTRG method, thermodynamic functions, such as the magnetization 
per site $m( T )$ and the internal energy per bond $u( T )$, can be calculated from the free energy per
site $f^{(\infty)}_{~}$. Alternatively, these functions are obtained by inserting
impurity tensors (separately derived from $C^{(n)}_{~}$ and $X^{(n)}_{~}$) into the tensor network
of the entire system. Since the fractal lattice under consideration is inhomogeneous, these 
thermodynamic functions can depend on the position they are placed. In order to check the dependence, 
we choose three typical locations $A$, $B$, and $Y$, as shown in Fig.~\ref{fig:Fig_5} on the fractal lattice. 

As an example of such a single site function, let us consider a tensor representation of the local 
magnetization. Looking at the position of site $A$ in Fig.~\ref{fig:Fig_5}, one finds that it is
located on the corner matrix $C^{(1)}_{~}$. 
Thus, the initial impurity tensor on that location is expressed as
\begin{equation}
A_{ij}^{(1)} = \sum_{\xi = \pm 1}^{~} \xi \, \exp{\bigl[ K \, \xi \left(\sigma_i^{~} + \sigma_j^{~} \right) \bigr]} \, ,
\label{Eq_13}
\end{equation}
similar to Eq.~\eqref{Eq_3}. It is also easy to check that the initial impurity tensor $B^{(1)}_{~}$, which 
is placed on a position different from $A$, is expressed by the identical equation, so that we have  
$A_{ij}^{(1)} = B_{ij}^{(1)}$. The site $Y$ lies inside the area $X^{(1)}_{~}$ and we define the corresponding 
initial tensor for local magnetization as
\begin{align}
Y_{ijkl}^{(1)} = & \sum_{\xi\eta}^{~} \frac{\xi + \eta}{2}
\exp{\bigl[K \, ( \sigma_i^{~} \sigma_j^{~} + \sigma_k^{~} \sigma_l^{~} + \xi\eta) \bigr]} 
\nonumber\\ 
&\times \exp{\bigl[K \, \xi \left( \sigma_j^{~} + \sigma_k^{~} \right)
+ K \, \eta \left( \sigma_i^{~} + \sigma_l^{~} \right) \bigr]} \, ,
\label{Eq_14}
\end{align}
similarly to Eq.~\eqref{Eq_6}. 

\begin{figure}[tb]
\includegraphics[width=0.45 \textwidth]{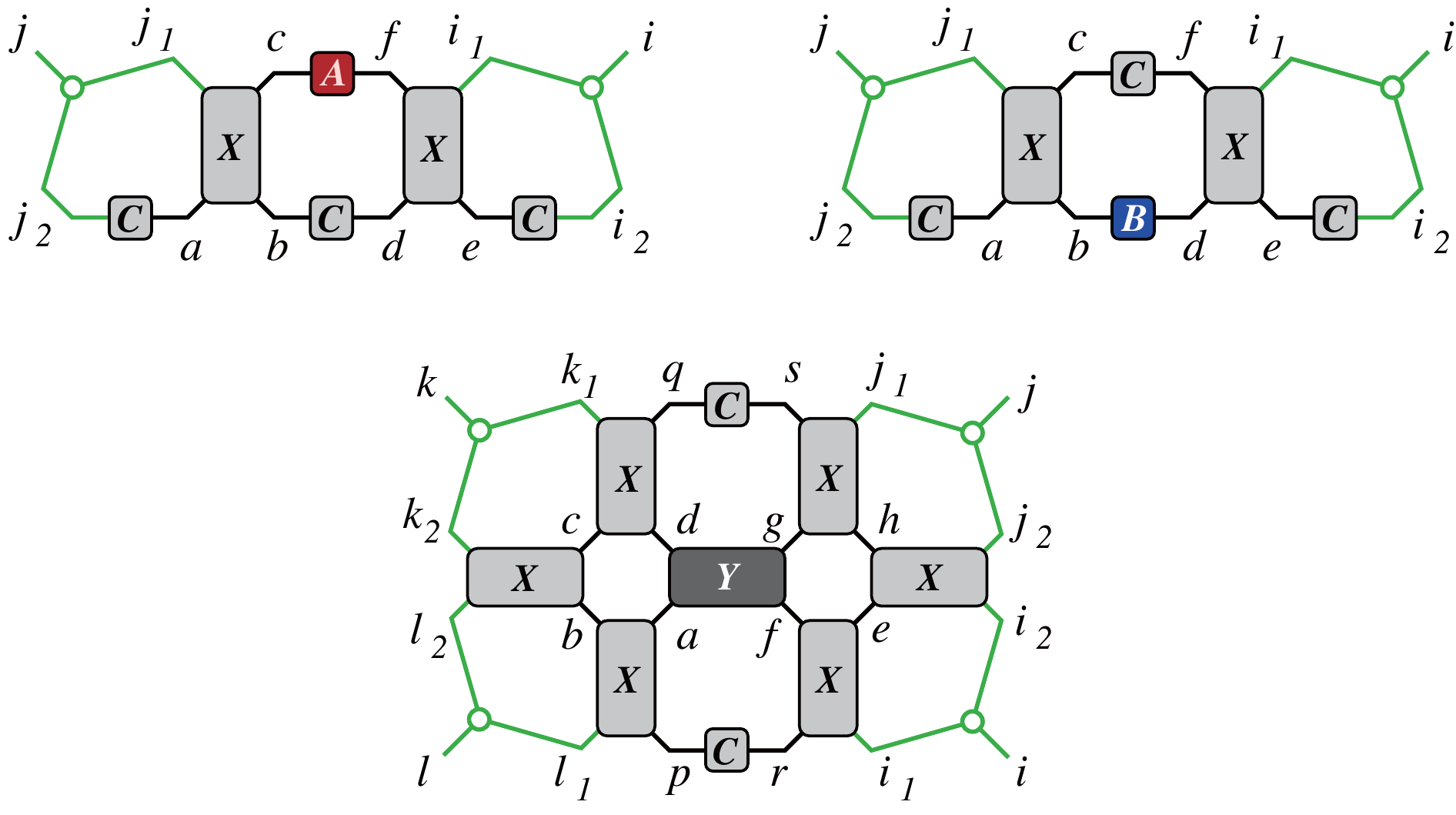}
\caption{(Color online)
Extension processes of the impurity tensor in Eq.~\eqref{Eq_15} (top left), Eq.~\eqref{Eq_16} (top right), 
and Eq.~\eqref{Eq_17} (bottom). The RG transformations $U$ are expressed by the external
three green lines meeting at the open circles.
}
\label{fig:Fig_6}
\end{figure}

We can thus build up analogous extension processes of tensors, each of which contains an impurity tensor we
have defined. The extension process of the impurity corner matrix that contains $A^{(1)}_{ij}$ is then written as
\begin{equation} 
A_{(i_1^{~} i_2^{~}) (j_1^{~} j_2^{~})}^{(n+1)} = \sum\limits_{abcdef}
C_{a j_2^{~}}^{(n)} X_{a b c j_1^{~}}^{(n)} A_{f c}^{(n)} C_{d b}^{(n)} X_{d e i_1^{~} f}^{(n)} C_{i_2^{~} e}^{(n)} \, , 
\label{Eq_15}
\end{equation}
which is graphically shown in Fig.~\ref{fig:Fig_6} (top left). Therein, the RG transformation
$A_{(i_1^{~} i_2^{~}) (j_1^{~} j_2^{~})}^{(n+1)} \to A_{ij}^{(n+1)}$
is depicted by the green lines with the open circles, which stand for $U$ in accord with Eq.~\eqref{Eq_12}. The impurity tensor placed
around the site $B$ obeys the extension procedure
\begin{equation}
B_{(i_1^{~} i_2^{~}) (j_1^{~} j_2^{~})}^{(n+1)} = \sum\limits_{abcdef}
C_{a j_2^{~}}^{(n)} X_{a b c j_1^{~}}^{(n)} C_{f c}^{(n)} B_{d b}^{(n)} X_{d e i_1^{~} f}^{(n)} C_{i_2^{~} e}^{(n)} \, , 
\label{Eq_16}
\end{equation}
as shown on the top right of Fig.~\ref{fig:Fig_6} (top right). For the location $Y$ shown in Fig.~\ref{fig:Fig_5},
we take the contraction
\begin{equation} 
\begin{split}
Y_{(i_1^{~} i_2^{~}) (j_1^{~} j_2^{~}) (k_1^{~} k_2^{~}) (l_1^{~} l_2^{~}) }^{(n+1)} = \sum\limits_{{a b c d e f}\atop{g h p r q s}}
X_{a b l_1^{~} p}^{(n)} X_{b c k_2^{~} l_2}^{(n)} X_{c d q k_1^{~}}^{(n)} \\
Y_{f g d a}^{(n)} X_{e f r i_1^{~}}^{(n)} X_{g h j_1^{~} s}^{(n)} 
X_{i_2^{~} j_2^{~} h e}^{(n)} C_{r p}^{(n)} C_{s q}^{(n)}  \, , 
\end{split}
\label{Eq_17}
\end{equation}
which is depicted in Fig.~\ref{fig:Fig_6} (bottom), where the graph is rotated by the right angle for book keeping.  

\begin{figure}[tb]
\includegraphics[width=0.48 \textwidth]{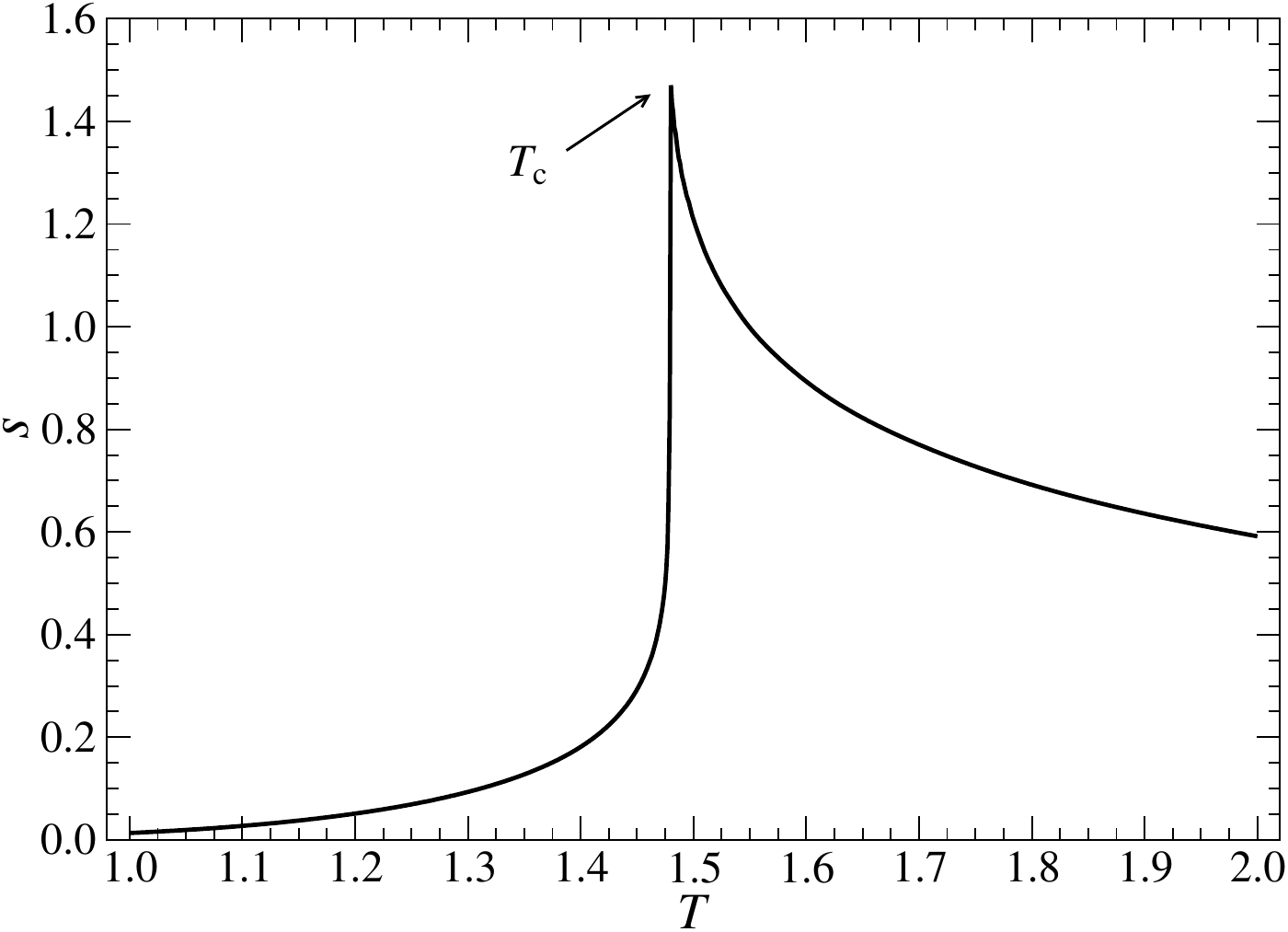}
\caption{The entanglement entropy $s( T )$ in Eq.~\eqref{eq:entropy}.}
\label{fig:Fig_7}
\end{figure}

In the calculation of the local bond energy $u( T )$, the initial tensors satisfy the equations
\begin{align}
A_{ij}^{(1)} &= - \frac{J}{2} \left( \sigma_i^{~} + \sigma_j^{~} \right) 
\sum_{\xi}^{~} \xi \, \exp{\bigl[ K \xi \left( \sigma_i^{~} + \sigma_j^{~} \right) \bigr]} \, , \\
\begin{split}
Y_{ijkl}^{(1)} &= \sum_{\xi\eta} - J \xi\eta \, 
\exp{\bigl[ K( \sigma_i^{~} \sigma_j^{~} + \sigma_k^{~} \sigma_l^{~} + \xi\eta) \bigr]} \\ 
&\times \exp{\bigl[ K \xi \left(\sigma_j^{~} + \sigma_k^{~} \right)
+ \eta \left(\sigma_i^{~} + \sigma_l^{~} \right) \bigr]} \, ,
\end{split}
\end{align}
recalling that $B_{ij}^{(1)} = A_{ij}^{(1)}$. Starting the extension processes with these 
initial tensors, we can calculate the expectation value of the bond energy around the site $A$ by 
means of the ratio
\begin{equation}
u_{\rm A}^{~}( T ) = \lim_{n \rightarrow \infty}^{~}
\frac{{\rm Tr} \, \Bigl( A^{(n)}_{~} \left[ C^{(n)}_{~} \right]^3 \Bigr)}{{\rm Tr}\, \Bigl( \left[ C^{(n)}_{~} \right]^4 \Bigr)} \, .
\end{equation}
The convergence with respect to $n$ is fast because of the fractal geometry.
It is straightforward to obtain the local energy $u_{\rm B}^{~}( T )$ and $u_{\rm Y}^{~}( T )$, as well as 
the local magnetization $m_{\rm A}^{~}( T )$, $m_{\rm B}^{~}( T )$, and $m_{\rm Y}^{~}( T )$ in the
same manner.

\section{Numerical Results}

For simplicity, we use the temperature scale with 
$k_{\rm B}^{~} = 1$ and fix the ferromagnetic interaction strength to $J = 1$. All the shown
data are obtained after taking a sufficiently large number of system extensions, provided that the convergence
with respect to $n$ has been reached. The degrees of freedom $D$ for each leg-dimension is $D = 28$ at most. 
Apart from the critical (phase transition) region, where $D$ needs to be the largest, we used $D = 18$, which
sufficed to obtain precise and converged data we have used for drawing all the graphs.

An analogous kind of the entanglement entropy $s( T )$ can be calculated by the HOTRG method. After applying 
SVD to the extended tensor, $s( T )$ can be naturally obtained from the singular values $\omega_{\xi}^{~}$
in Eq.~\eqref{Eq_10} through the formula
\begin{equation} \label{eq:entropy}
s( T ) = -\sum_{\xi}^{~} \frac{\omega_{\xi}^2}{\Omega} \, \ln \frac{\omega_{\xi}^2}{\Omega}  \, , 
\end{equation}
where $\Omega = \sum_{\xi}^{~} \omega_{\xi}^2$ normalizes the probability. The entanglement entropy 
$s( T )$ always exhibits stable convergence with respect to $n$. Figure \ref{fig:Fig_7} shows the temperature 
dependence of $s( T )$, which is obtained with $D = 18$. There is a sharp peak at the critical temperature 
$T_{\rm c}^{~}$, which can be roughly determined as $1.48$ from the data shown.

\begin{figure}[tb]
\includegraphics[width=0.48 \textwidth]{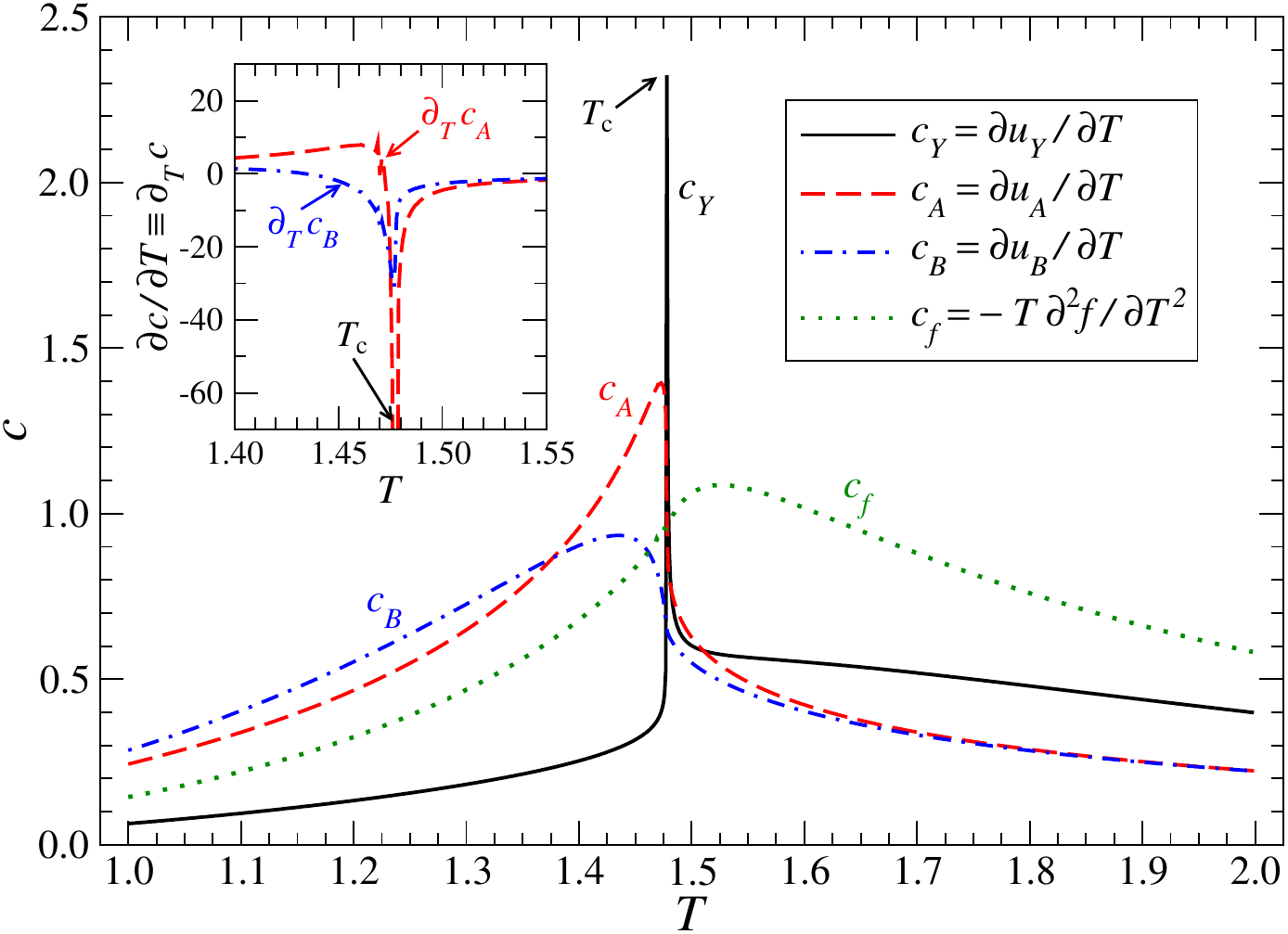}
\caption{ (Color online)
Specific heats $c_A^{~}( T )$, $c_B^{~}( T )$, $c_Y^{~}( T )$, and $c_f^{~}( T )$. The inset shows
the derivative of the specific heat with respect to temperature $\partial_T c$.}
\label{fig:Fig_8}
\end{figure}

Taking the numerical derivative with respect to $T$ for the calculated local energies $u_{\rm A}^{~}( T )$, 
$u_{\rm B}^{~}( T )$, and $u_{\rm Y}^{~}( T )$, respectively, we obtain the specific heats $c_{\rm A}^{~}( T )$, 
$c_{\rm B}^{~}( T )$, and $c_{\rm Y}^{~}( T )$, as shown Fig.~8. We observe a sharp peak in $c_{\rm Y}^{~}( T )$ 
at $T_{\rm c}^{~}$, whereas there is only a rounded maximum in $c_{\rm A}^{~}( T )$ and $c_{\rm B}^{~}( T )$, 
and their peak positions do not coincide with $T_{\rm c}^{~}$ associated with the position $Y$. The specific 
heat per site $c_f^{~}( T )$ defined in Sec.~II as well as $c_{\rm A}^{~}( T )$ and $c_{\rm B}^{~}( T )$ demonstrate 
a weak singularity at $T_{\rm c}^{~}$. This fact can be confirmed by taking their derivative with respect to $T$,
i.e., $\frac{\partial c}{\partial T}$, which leads to the identical singularity at $T_{\rm c}^{~}$, as shown in the inset
of Fig.~\ref{fig:Fig_8}. The result clearly manifests that the critical behavior strongly depends on the location,
where the measurements of the bond energy is carried out.

\begin{figure}[tb]
\includegraphics[width=0.48 \textwidth]{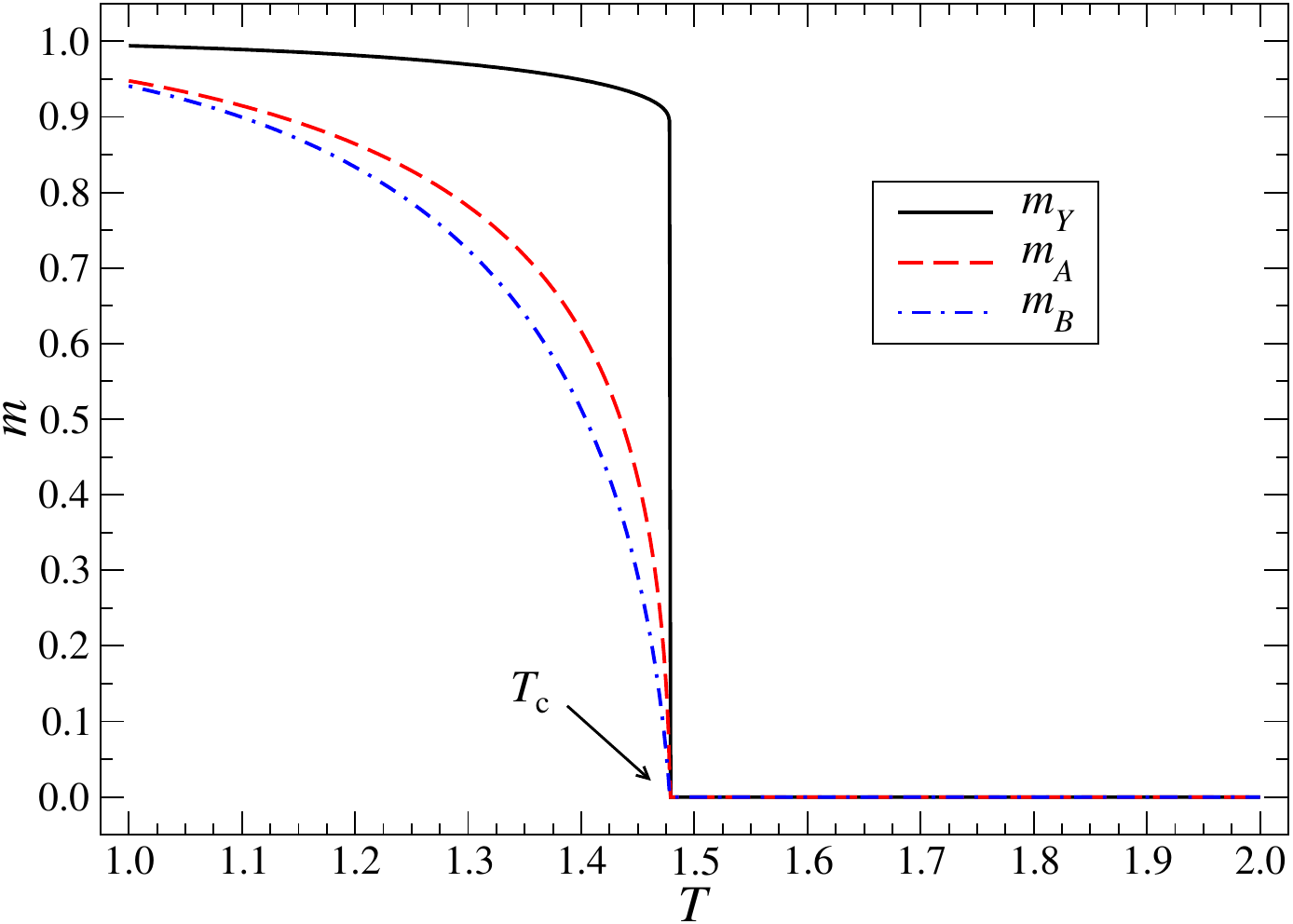}
\caption{ (Color online)
The local magnetization $m_{\rm Y}^{~}( T )$, $m_{\rm A}^{~}( T )$, and $m_{\rm B}^{~}( T )$. 
}
\label{fig:Fig_9}
\end{figure}

\begin{figure}[tb]
\includegraphics[width=0.48 \textwidth]{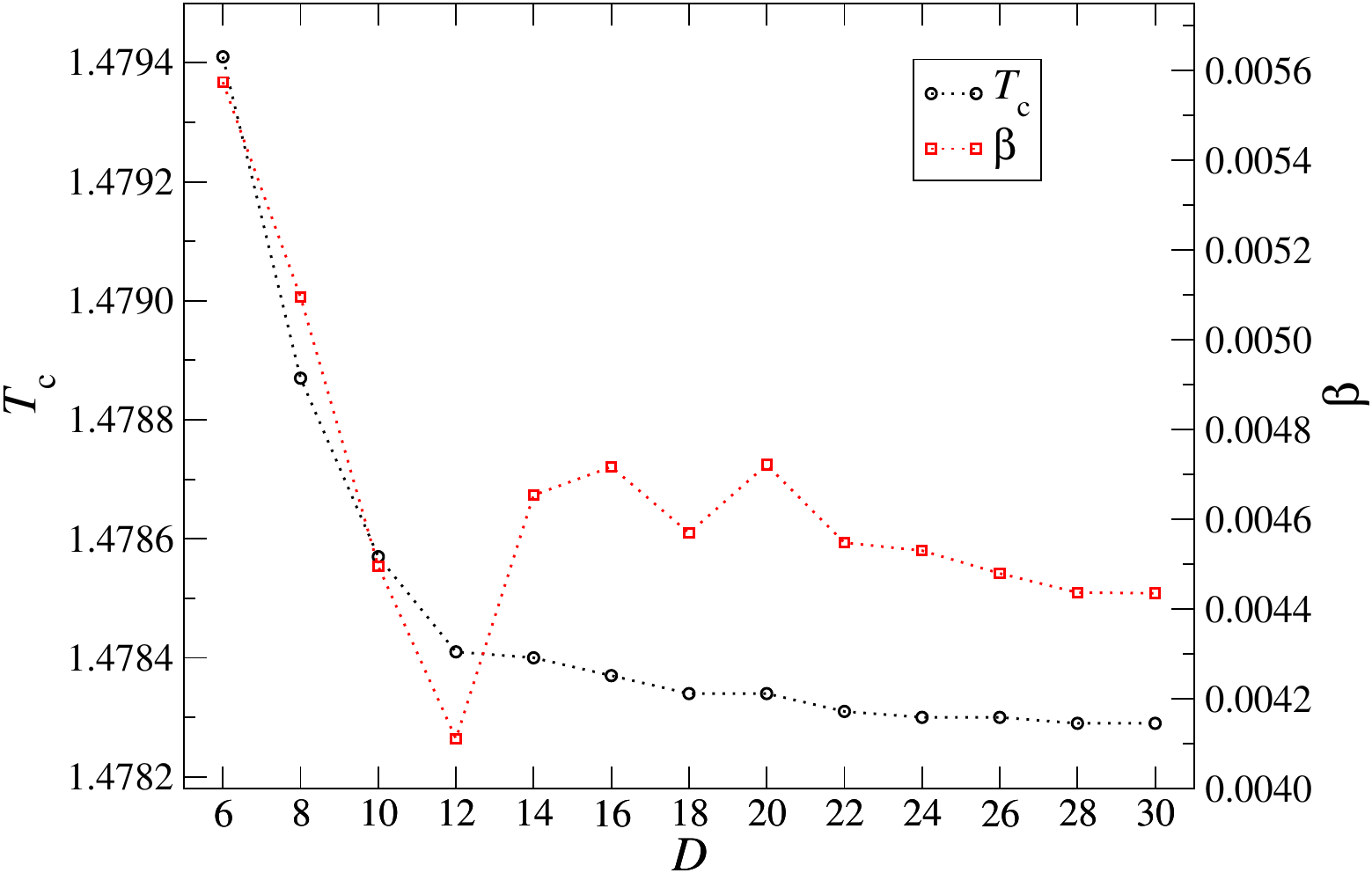}
\caption{ (Color online)
Critical temperature $T_{\rm c}^{~}$ (black circles) and critical exponent $\beta$ (red squares) with respect to the bond dimension $D$ obtained from the local magnetization $m_{\rm Y}^{~}( T )$.  
}
\label{fig:Fig_10}
\end{figure}

\begin{figure}[tb]
\includegraphics[width=0.48 \textwidth]{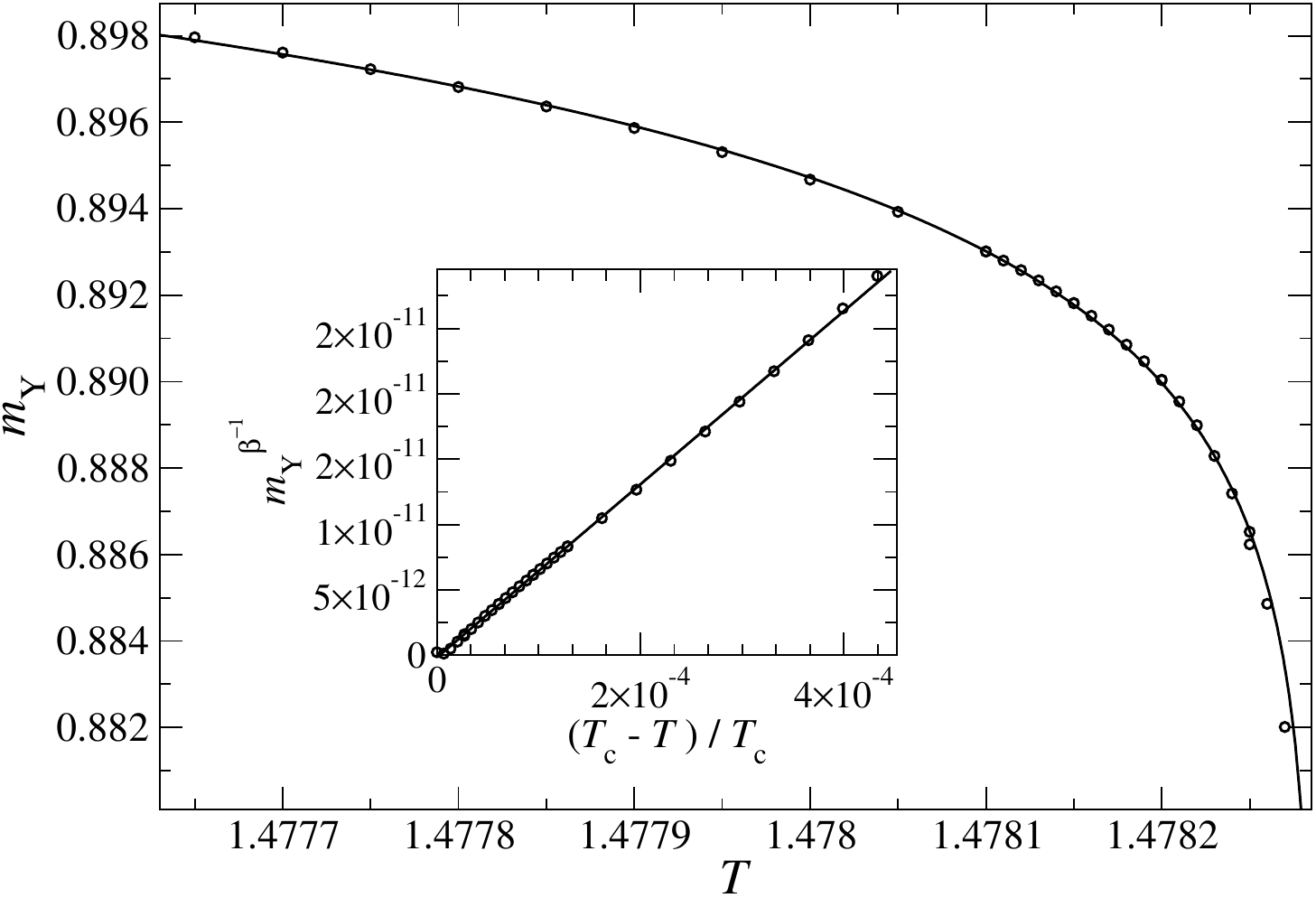}
\caption{
Detailed view of $m_{\rm Y}^{~}( T )$ when $D = 30$. 
Inset: the power-law behavior below $T_{\rm c}^{~} =1.47829$ plotted with the exponent $\beta = 0.0044(4)$.  
}
\label{fig:Fig_11}
\end{figure}

Figure~\ref{fig:Fig_9} shows the local magnetizations $m_{\rm A}^{~}( T )$, $m_{\rm B}^{~}( T )$, and 
$m_{\rm Y}^{~}( T )$ with respect to temperature $T$, under the condition that $D = 18$. They fall to zero 
simultaneously at the identical $T_{\rm c}^{~}$, while the critical exponent $\beta$ in 
$m( T ) \propto ( T_{\rm c}^{~} - T )^{\beta}_{~}$ is significantly different for each case. 
From the plotted $m_{\rm A}^{~}( T )$ we obtain $\beta \approx 0.52$, and
from $m_{\rm B}^{~}( T )$ we obtain $\beta \approx 0.78$. 
In both cases we use the rough estimate $T_{\rm c}^{~} \approx 1.478$, and the
data in the range $| T_{\rm c}^{~} - T | < 0.015$ are considered for numerical fitting. 
Since the variation in $m_{\rm Y}^{~}( T )$ is too rapid to capture $\beta$ under the condition $D = 18$, 
we increase the tensor-leg freedom up to $D = 30$. 
As can be seen in Figure~\ref{fig:Fig_10}, both critical temperature $T_{\rm c}^{~}$ and exponent $\beta$ obtained from the local magnetization $m_{\rm Y}^{~}( T )$ appear to be well converged when $D\gtrsim28$. 
Figure~\ref{fig:Fig_11} shows $m_{\rm Y}^{~}( T )$ 
zoomed-in around $T \sim 1.478$. It should be noted that a small numerical error is strongly amplified in the 
temperature region $| T  -  T_{\rm c}^{~} | \lesssim 10^{-5}_{~}$. Therefore, the data points in this narrow 
region were excluded from the fitting analysis. Then, we obtain $T_{\rm c}^{~} \approx 1.47829$. 
The estimated critical exponent $\beta = 0.0044(4)$ is roughly two orders of magnitude smaller than
$\beta$ obtained from $m_{\rm A}^{~}( T )$ and $m_{\rm B}^{~}( T )$. In a similar manner, as we have observed for the specific heat, 
the critical behavior of the model strongly depends on the location of the impurity tensors $A$, $B$, and $Y$ on the Sierpi\'{n}ski carpet.

\begin{figure}[tb]
\includegraphics[width=0.48 \textwidth]{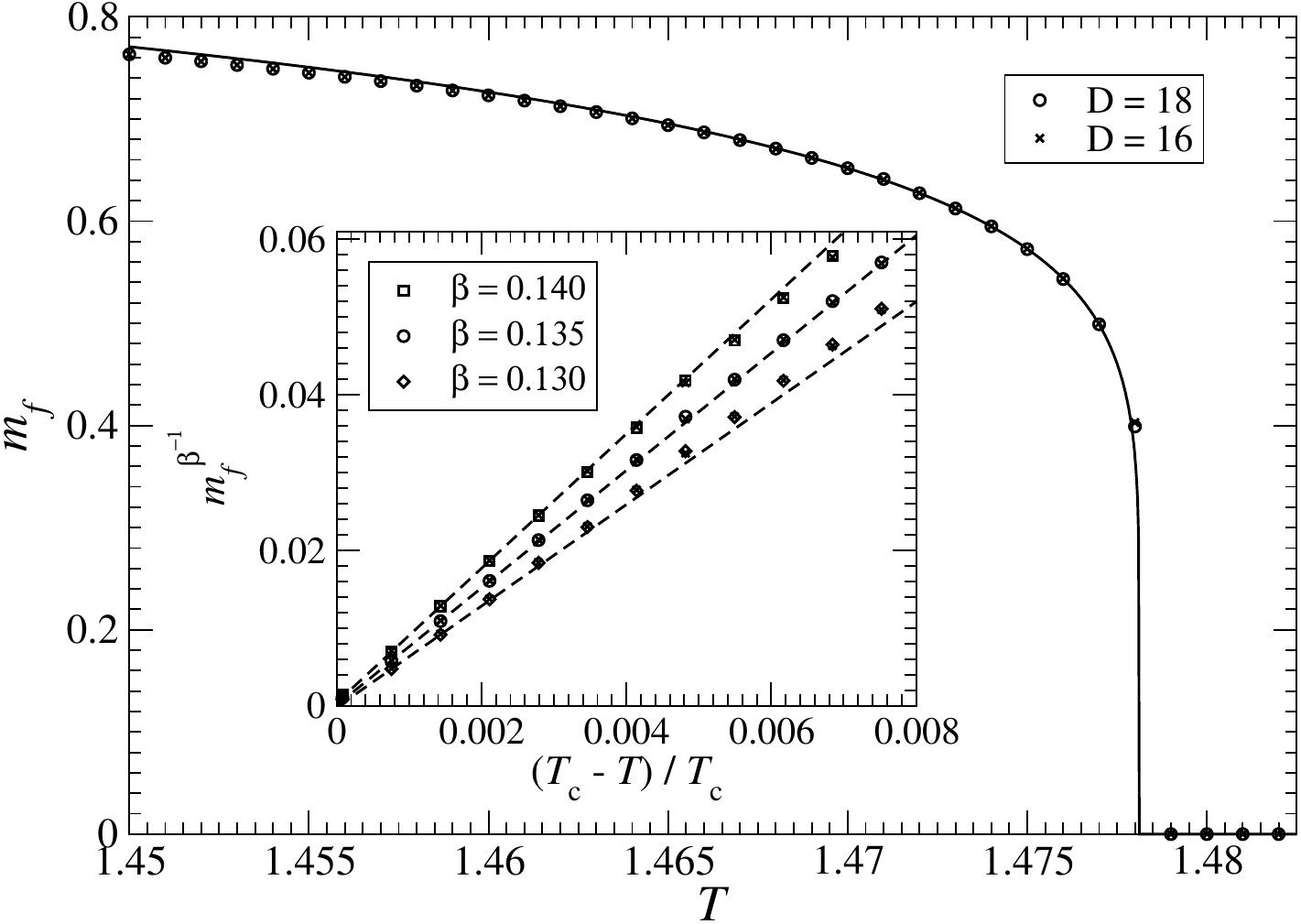}
\caption{
%
The global magnetization $m_f(T)$ when $D = 16$ and $D = 18$. 
Inset: the power-law behavior below $T_{\rm c}^{~}$.  
%
}
\label{fig:Fig12}
\end{figure}

%
The global spontaneous magnetization $m_f$ calculated according to the Eq.~\eqref{Eq:glob_mag} using automatic differentiation is presented in Fig.~\ref{fig:Fig12}. 
For $D = 18$, the fitting of $m_{f}(T)$ in the region $T \lesssim T_{\rm c}^{~}$ yields $T_{\rm c}^{~} \approx 1.478$ and $\beta \approx 0.135$. 
Relative difference between $D = 16$ and $D = 18$ in the estimate of $\beta$ is around $0.1\%$.
The linear dependence (dashed lines) of $m^{\beta^{-1}}_{f}$ below $T_{\rm c}^{~}$ is shown in the inset of Fig.~\ref{fig:Fig12}.

\section{Conclusions and Discussions}

We have investigated the phase transition of the ferromagnetic Ising model on the Sierpi\'{n}ski 
carpet. The numerical procedures in the HOTRG method are modified, so that they fit the 
recursive structure in the fractal lattice. We have confirmed the presence of the 
second order phase transition, which is located around $T_{\rm c}^{~} \approx {1.478}$, in 
accordance with the previous studies~\cite{Carmona, Monceau1, Monceau2, Pruessner, Bab, Bab2}. 
Notably, the HOTRG method used in this study achieved numerical convergence of physical observables with $k \sim 35$ iterative extensions (generations) of the system, which is significantly higher than the maximum value of $k \leq 8$ reached by Monte Carlo studies. 
This demonstrates the effectiveness of the HOTRG method for studying phase transitions on fractal lattices.
Moreover, the global behavior of the entire system captured by the free energy per site $f^{(\infty)}_{~}$ 
exhibits the presence of a very weak singularity at $T_{\rm c}^{~}$, as we observed in Ref.~\onlinecite{2dising}.

What is characteristic of this fractal lattice is the position dependence in the local magnetization
$m( T )$ and local energy $u( T )$. For example, we find that the critical exponent $\beta$ differs by 
a couple of orders of magnitude, which corresponds to the fact that the measured magnetization depends on 
position, where the impurity tensor is placed on the fractal-lattice. A key feature appears in the local energy
$u_{\rm Y}^{~}( T )$, where we deduce a sharp peak in its temperature derivative, $c_{\rm Y}^{~}( T )$; 
contrary to the smooth behavior in $c_f^{~}( T )$, being the averaged specific heat. Intuitively, such 
position dependence would be explained by the 
density of sites around the pinpointed location. Around the site Y, the spins are interconnected more densely 
than those around the boundary sites A and B in Fig.~\ref{fig:Fig_5}. One might find a similarity with the critical
behavior on the Bethe lattice\cite{Baxter,p4hyper}, where the singular behavior is only visible deep inside the system,
whereas the free energy is represented by an analytic function of $T$ for the entire lattice. 
Lastly, let us also mention that the position dependence of the critical exponents we observed in this study is analogous to the surface critical behavior captured by boundary tensor network methods in Refs.~\onlinecite{btrg, btnr}.

%
Finally, we leveraged automatic differentiation to compute the global spontaneous magnetization $m_f(T)$, which represents the average magnetization over all site locations. 
This approach allowed us to overcome the challenges associated with averaging impurities over all site locations on the fractal lattice. 
Our analysis revealed that the associated global critical exponent $\beta \approx 0.135$, which is intermediate between the local exponents associated with $m_A$ ($\beta \approx 0.52$) and $m_B$ ($\beta \approx 0.78$) on the one hand and $m_Y$ ($\beta \approx 0.004$) on the other hand, but much closer to the former. 
Notably, the global critical exponent we report in this study is consistent with estimates from previous Monte Carlo studies, as reported in the literature~\cite{Carmona, Monceau1, Monceau2, Bab, Bab2, MCRG}. 
Our findings have important implications for understanding the critical behavior of magnetic systems on fractal lattices and could guide future experimental and theoretical investigations.
%

The current study can be extended to other fractal lattices, e.g., variants of the Sierpi\'{n}ski carpet or a fractal lattice we had already studied earlier\cite{2dising}, where the positional dependence of the impurities has not been examined yet. Another point to consider is to investigate more variations of the locations on the fractal lattice to analyze the mechanism of the non-trivial position-dependent behavior we observed.

\begin{acknowledgments}
This work was partially funded by Agent\'{u}ra pre Podporu V\'{y}skumu a V\'{y}voja (No. APVV-20-0150),
Vedeck\'{a} Grantov\'{a} Agent\'{u}ra M\v{S}VVa\v{S} SR a SAV (VEGA Grant No. 2/0156/22) and Joint Research
Project SAS-MOST 108-2112-M-002-020-MY3.
T.N. and A.G. acknowledge the support of Ministry of Education, Culture, Sports, Science and Technology 
(Grant-in-Aid for Scientific Research JSPS KAKENHI 17K05578, 17F17750 and 21K03403). 
J.G. was supported by Japan Society for the Promotion of Science (P17750). 
J.G. also acknowledges support from the National Science and Technology Council, the Ministry of Education (Higher Education Sprout Project NTU-111L104022), and the National Center for Theoretical Sciences of Taiwan.
T.N. thanks the funding of the Grant-in-Aid for Scientific Research MEXT ``Exploratory Challenge on Post-K computer'' (Frontiers of Basic Science: Challenging the Limits). 
\end{acknowledgments}

\appendix

\section{Efficient Model Representation and RG Transformation} \label{app_A}

Here, we define a computationally more efficient tensor network representation than one introduced in the main text. 
This is achieved by replacing the corner matrix $C$ by its two halves and re-expressing the order-four tensor $X$ in terms of two order-three tensors. Using this approach, the overall computational cost is reduced from ${\cal O}(D^{10})$ to ${\cal O} (D^8)$ and the memory cost is reduced from ${\cal O} (D^8)$ to ${\cal O} (D^6)$, where $D$ is the bond dimension cutoff. 

\begin{figure}[tb]
\includegraphics[width=0.26 \textwidth]{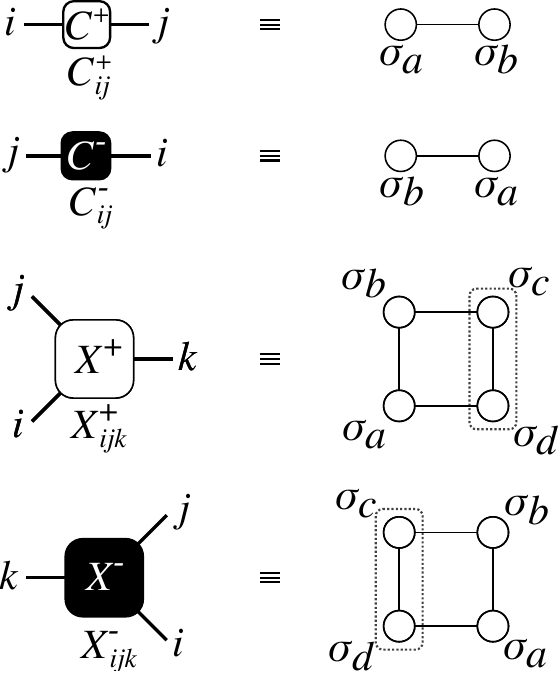}
\caption{
Structure of the initial half-corner matrices $C_{~}^{(1), +}$ and $C_{~}^{(1), -}$ in Eqs.~\eqref{Eq_A1}-\eqref{Eq_A2} and two 3-leg tensors $X_{ijk}^{(1), + }$ and $X_{ijk}^{(1), - }$ in Eqs.~\eqref{Eq_A3}-\eqref{Eq_A4}.}
\label{fig:Fig_13}
\end{figure}

\subsubsection*{Initialization} At initial step $n=1$, we define ``left'' and ``right'' halves of the corner matrix (\textit{cf.} Eq.~\eqref{Eq_3})
\begin{equation} \label{Eq_A1}
C_{ij}^{(1), + } = \exp\bigl[ K \left( \sigma_a^{~} + \sigma_b^{~} \right) \bigr] \, ~
\end{equation}
and
\begin{equation} \label{Eq_A2}
C_{ij}^{(1), - } = \exp\bigl[ K \left( \sigma_b^{~} + \sigma_a^{~} \right) \bigr] \, ,
\end{equation}
respectively, where the matrix indices $i = ( \sigma_a^{~} + 1 ) / 2$ and $j = ( \sigma_b^{~} + 1 ) / 2$ take the value either $0$ or $1$. Notice that the (``left'' half) $C_{~}^{+}$ is indexed left-to-right whereas the (``right'' half) $C_{~}^{-}$ is indexed right-to-left, as seen in Fig.~\ref{fig:Fig_13}. 

At the same time, we initialize the two halves (i.~e., order-three tensors) of the $X$ region (i.~e., order-four tensor) as follows (\textit{cf.} Eq.~\eqref{Eq_6})  
\begin{equation} \label{Eq_A3}
X_{ijk}^{(1), + } = \exp\bigl[ K \left( \sigma_a^{~} \sigma_b^{~} + 
\sigma_b^{~} \sigma_c^{~} + \sigma_c^{~} \sigma_d^{~} + \sigma_d^{~} \sigma_a^{~} \right)  \bigr]  \, ,
\end{equation}
and
\begin{equation} \label{Eq_A4}
X_{ijk}^{(1), - } = \exp\bigl[ K \left( \sigma_d^{~} \sigma_c^{~} + 
\sigma_c^{~} \sigma_b^{~} + \sigma_b^{~} \sigma_a^{~} + \sigma_a^{~} \sigma_d^{~} \right)  \bigr]  \, ,
\end{equation}
where $k$ is a combined index obtained from $c$ and $d$, i.~e., $k = \sigma_c^{~} + 1 + ( \sigma_d^{~} + 1 ) / 2$, and it takes four integer values ($0$ to $3$). Notice that $X_{~}^{+}$ is indexed clockwise, whereas $X_{~}^{-}$ is indexed anti-clockwise, see Fig.~\ref{fig:Fig_13}. 

\begin{figure}[tb]
\includegraphics[width=0.45 \textwidth]{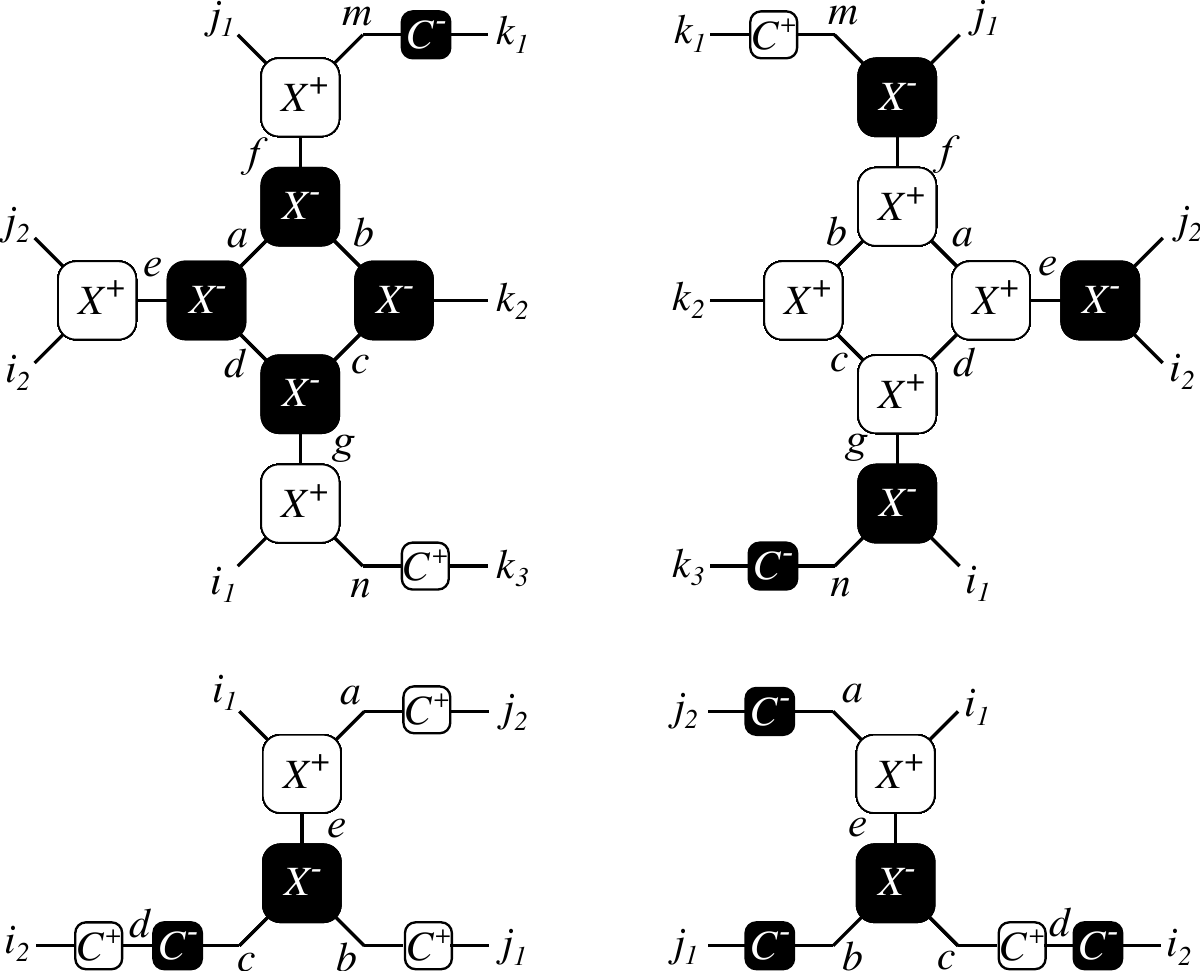}
\caption{
The extension of the half-corner matrices $C^{(n), +}_{~}$ (lower left) and $C^{(n), -}_{~}$ (lower right) according to Eq.~\eqref{Eq_A5} and Eq.~\eqref{Eq_A6}, respectively, and the tensors $X_{~}^{(n+1), + }$ (upper left) and $X_{~}^{(n+1), - }$ (upper right) according to Eq.~\eqref{Eq_A7} and Eq.~\eqref{Eq_A8}, respectively.}
\label{fig:Fig_14}
\end{figure}

\subsubsection*{Extensions}

The extended corner-matrix halves $C_{~}^{(n+1), + }$ and $C_{~}^{(n+1), - }$ are 
\begin{equation}  \label{Eq_A5}
\begin{split}
&C_{ij}^{(n+1), + } = C_{( i_1^{~} i_2^{~}) (j_1^{~} j_2^{~})}^{(n+1), + } = \\ 
&\sum_{a b c d e}^{~} 
X_{i_1 a e}^{(n), + } 
X_{c b e}^{(n), - }
C_{i_2 d}^{(n), + } 
C_{a j_2}^{(n), + } 
C_{b j_1}^{(n), + }  
C_{c d}^{(n), - } \, ,
\end{split} 
\end{equation}
and
\begin{equation}  \label{Eq_A6}
\begin{split}
&C_{ij}^{(n+1), - } = C_{( i_1^{~} i_2^{~}) (j_1^{~} j_2^{~})}^{(n+1), - } = \\ 
&\sum_{a b c d e}^{~} 
X_{b c e}^{(n), - } 
X_{a i_1 e}^{(n), + }
C_{a j_2}^{(n), -} 
C_{b j_1}^{(n), - } 
C_{i_2 d}^{(n), - }  
C_{c d}^{(n), + } \, ,
\end{split} 
\end{equation}
respectively, where the new indices $i$ and $j$, represent the grouped indices $( i_1^{~} i_2^{~})$ and $(j_1^{~} j_2^{~})$, respectively, see lower row of Fig.~\ref{fig:Fig_14}.

Similarly, the extension relations for $X^{(n+1), + }_{~}$ and $X^{(n+1), - }_{~}$ are
\begin{equation}  \label{Eq_A7}
\begin{split}
X_{i j k}^{(n+1), + } & = X_{( i_1^{~} i_2^{~}) (j_1^{~} j_2^{~}) ( k_1^{~} k_2^{~} k_3^{~})}^{(n+1), + } \\
& =\sum\limits_{{a b c d e f}\atop{g m n}} 
X_{n i_1^{~} g}^{(n), + } X_{i_2^{~} j_2^{~} e}^{(n), + } 
X_{j_1^{~} m f}^{(n), + } \\
& \hspace{1.4cm} X_{d a e}^{(n), - } X_{a b f}^{(n), - } 
X_{b c k_2^{~}}^{(n), - } X_{c d g}^{(n), - } \\ 
& \hspace{1.4cm} C_{n k_3^{~}}^{(n), + } C_{m k_1^{~}}^{(n), - } \, , \end{split} 
\end{equation}
\begin{equation}  \label{Eq_A8}
\begin{split}
X_{i j k}^{(n+1), - } & = X_{( i_1^{~} i_2^{~}) (j_1^{~} j_2^{~}) ( k_1^{~} k_2^{~} k_3^{~})}^{(n+1), - } \\
& =\sum\limits_{{a b c d e f}\atop{g m n}} 
X_{d a e}^{(n), + } X_{c d g}^{(n), + } 
X_{b c k_2^{~}}^{(n), + } X_{a b f}^{(n), +} \\
& \hspace{1.4cm} X_{n i_1^{~} g}^{(n), - } X_{i_2^{~} j_2^{~} e}^{(n), - } 
X_{j_1^{~} m f}^{(n), - } \\ 
& \hspace{1.4cm} C_{m k_1^{~}}^{(n), + } C_{n k_3^{~}}^{(n), - } \, , \end{split} 
\end{equation}
where we have abbreviated the grouped indices to $i = (i_1^{~} i_2^{~})$, $j = (j_1^{~} j_2^{~})$, 
$k = (k_1^{~} k_2^{~})$, and $l = (l_1^{~} l_2^{~})$, see the upper row of Fig.~\ref{fig:Fig_14}.

\begin{figure}[tb]
\includegraphics[width=0.30 \textwidth]{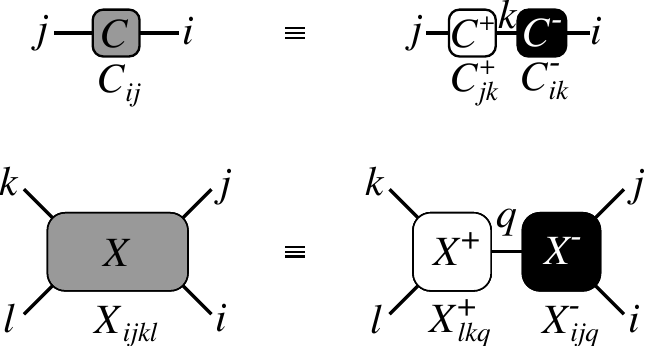}
\caption{
	Upper row: Mapping between the corner matrix $C$ and its halves $C^+$ and $C^-$ in Eq.~\eqref{Eq_A9}.
	Lower row: Mapping the tensor $X$ and its halves $X^+$ and $X^-$ (lower row) in Eq.~\eqref{Eq_A10}.}
\label{fig:Fig_15}
\end{figure}

\subsubsection*{Mapping to the original model representation}
The full corner matrix $C^{(n)}$ and the tensor $X^{(n)}$ can be obtained at each step $n$ of the iterative extension process in a straightforward way 
\begin{equation}  \label{Eq_A9}
C_{ij}^{(n)} = \displaystyle \sum_{k} C_{jk}^{(n), + } C_{ki}^{(n), - } \, ~
\end{equation}
and
\begin{equation}  \label{Eq_A10}
X_{i j k l}^{(n)} = \displaystyle \sum_{q} X_{l k q}^{(n), + } X_{i j q}^{(n), - } \, ,
\end{equation}
respectively. This relation is depicted in Fig.~\ref{fig:Fig_15}. 

\subsubsection*{RG Transformation}
First, we reshape the extended tensor $X_{( i_1^{~} i_2^{~}) (j_1^{~} j_2^{~}) ( k_1^{~} k_2^{~} k_3^{~})}^{(n+1), + }$ into a matrix $\tilde{X}_{( i_1^{~} i_2^{~}) (j_1^{~} j_2^{~} k_1^{~} k_2^{~} k_3^{~})}^{(n+1), + }$. The projector $U$ is calculated using the singular value decomposition applied to $\tilde{X}_{~}^{(n+1), + }$
\begin{equation}  \label{Eq_A11}
\tilde{X}_{( i_1^{~} i_2^{~}) (j_1^{~} j_2^{~} k_1^{~} k_2^{~} k_3^{~})}^{(n+1), + } = 
\sum_{\xi=1}^{D} U_{(i_1^{~} i_2^{~}) \, \xi}^{~} \, \omega_{\xi}^{~} \, 
V^{~}_{(j_1^{~} j_2^{~} k_1^{~} k_2^{~} k_3^{~}) \, \xi } \, .
\end{equation}
To truncate the degrees of freedom associated with the grouped index $k=(k_1 k_2 k_3)$, we perform SVD on matrix $\tilde{\tilde{X}}_{( k_1^{~} k_2^{~} k_3^{~} ) (i_1^{~} i_2^{~} j_1^{~} j_2^{~} )}^{(n+1), + } = X_{( i_1^{~} i_2^{~}) (j_1^{~} j_2^{~}) ( k_1^{~} k_2^{~} k_3^{~})}^{(n+1), + }$ 
\begin{equation}  \label{Eq_A12}
\tilde{\tilde{X}}_{( k_1^{~} k_2^{~} k_3^{~} ) (i_1^{~} i_2^{~} j_1^{~} j_2^{~} )}^{(n+1), + } = 
\sum_{\xi=1}^{K_1} U_{(k_1^{~} k_2^{~} k_3^{~}) \, \xi}^{\prime} \, \omega_{\xi}^{\prime} \, 
V^{\prime}_{(i_1^{~} i_2^{~} j_1^{~} j_2^{~} ) \, \xi } \, ,
\end{equation}
where the sum corresponds to the largest $K_1$ singular values. 
\begin{figure}[th]
\includegraphics[width=0.45 \textwidth]{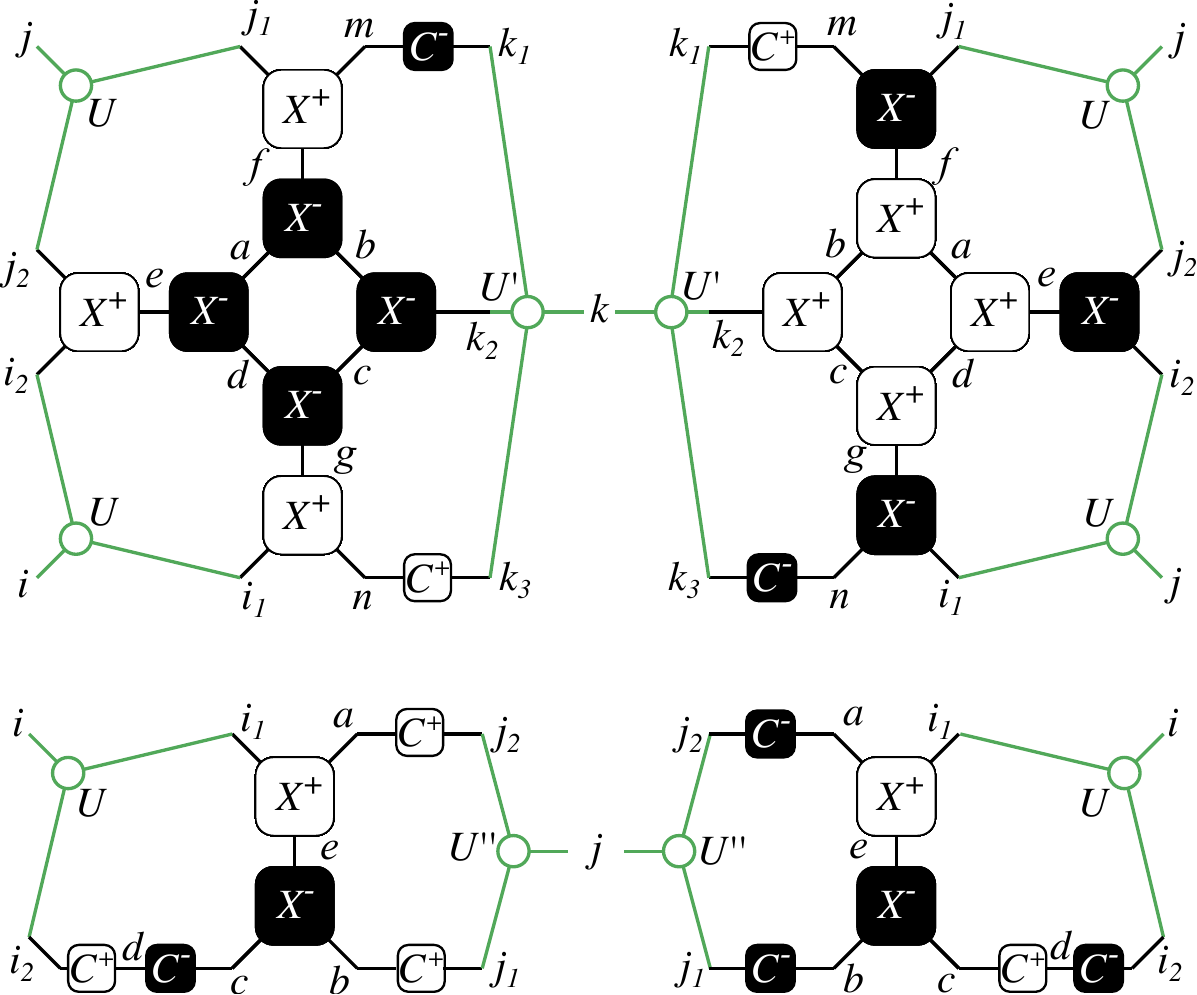}
\caption{(Color online) 
The renormalization group transformations in Eq.~\eqref{Eq_A14} (lower left), Eq.~\eqref{Eq_A15} (lower right), Eq.~\eqref{Eq_A16} (upper left), and  Eq.~\eqref{Eq_A17} (upper right).
}
\label{fig:Fig_16}
\end{figure}
Lastly, we prepare a projector for the grouped index $j=(j_1 j_2)$ in $\tilde{C}_{(j_1^{~} j_2^{~}) ( i_1^{~} i_2^{~})}^{(n+1), + } = C_{( i_1^{~} i_2^{~}) (j_1^{~} j_2^{~})}^{(n+1), + }$
\begin{equation}  \label{Eq_A13}
\tilde{C}_{(j_1^{~} j_2^{~}) ( i_1^{~} i_2^{~})}^{(n+1), + } = 
\sum_{\xi=1}^{K_2} U_{(j_1^{~} j_2^{~}) \, \xi}^{\prime \prime} \, \omega_{\xi}^{\prime \prime} \, 
V^{\prime \prime}_{(i_1^{~} i_2^{~} ) \, \xi } \, ,
\end{equation}
where the sum corresponds to the largest $K_2$ singular values. 

The RG transformation is then performed using the three different projectors $U^{~}_{~}$, $U^{\prime}_{~}$, and $U^{\prime \prime}_{~}$ as follows (see Fig.~\ref{fig:Fig_16})
\begin{equation} \label{Eq_A14}
C_{ij}^{(n+1), + } \leftarrow \sum_{i_1 i_2 j_1 j_2} U_{(i_1^{~} i_2^{~}) \, i}^{~} \, U_{(j_1^{~} j_2^{~}) \, j}^{\prime \prime} \,
C_{( i_1^{~} i_2^{~}) (j_1^{~} j_2^{~})}^{(n+1), + } \, ,
\end{equation}
\begin{equation}  \label{Eq_A15}
C_{ij}^{(n+1), - } \leftarrow \sum_{i_1 i_2 j_1 j_2} U_{(i_1^{~} i_2^{~}) \, i}^{~} \, U_{(j_1^{~} j_2^{~}) \, j}^{\prime \prime} \,
C_{( i_1^{~} i_2^{~}) (j_1^{~} j_2^{~})}^{(n+1), - } \, ,
\end{equation}
\begin{align}  \label{Eq_A16}
\begin{split}
X_{i j k}^{(n+1), + }  \leftarrow \sum_{{i_1 i_2 j_1 j_2}\atop{k_1 k_2 k_3}}
U_{(i_1^{~} i_2^{~}) \, i}^{~} & \, U_{(j_1^{~} j_2^{~}) \, j}^{~} \, 
U_{(k_1^{~} k_2^{~} k_3^{~}) \, k}^{\prime} \\
 & X_{(i_1^{~} i_2^{~}) (j_1^{~} j_2^{~}) (k_1^{~} k_2^{~} k_3^{~}) }^{(n+1), + } \, , 
\end{split}
\end{align}
\begin{align}  \label{Eq_A17}
\begin{split}
X_{i j k}^{(n+1), - }  \leftarrow \sum_{{i_1 i_2 j_1 j_2}\atop{k_1 k_2 k_3}}
U_{(i_1^{~} i_2^{~}) \, i}^{~} & \, U_{(j_1^{~} j_2^{~}) \, j}^{~} \, 
U_{(k_1^{~} k_2^{~} k_3^{~}) \, k}^{\prime} \\
 & X_{(i_1^{~} i_2^{~}) (j_1^{~} j_2^{~}) (k_1^{~} k_2^{~} k_3^{~}) }^{(n+1), - } \, . 
\end{split}
\end{align}

\end{document}